\RequirePackage{fix-cm}

\documentclass[twocolumn,epjc3]{svjour3}
\pdfoutput=1
\smartqed  % flush right qed marks, e.g. at end of proof

\RequirePackage{graphicx}

\usepackage{subfigure}
\usepackage{mathrsfs}
\usepackage[T1]{fontenc} % if needed
\usepackage{graphicx}
\usepackage{subfigure}
\usepackage{amsmath}
\usepackage{xspace}
\usepackage{url}
\usepackage{cite}
\usepackage{booktabs}
\usepackage{siunitx}
\usepackage[hyperindex,breaklinks]{hyperref}

\DeclareGraphicsExtensions{.pdf}

\usepackage{preprintcover}

\usepackage{lscape}
\newcommand*{\ytot}[1]{ \multicolumn{1}{r}{\num[round-mode=figures,round-precision=3]{#1}} }
\newcommand*{\yerr}[1]{ \multicolumn{1}{r}{$\pm$ \num[round-mode=figures,round-precision=2]{#1}} }

\newcommand{\wjetsabstract} {This paper presents cross sections for the production of a \Wg\ boson in association with jets, measured in proton--proton collisions at $\sqrt{s}=7\TeV$ with the ATLAS experiment at the Large Hadron Collider. With an integrated luminosity of $4.6\,\ifb$, this data set allows for an exploration of a large kinematic range, including jet production up to a transverse momentum of  $1 \TeV$ and multiplicities up to seven associated jets. The production cross sections for \Wg\ bosons are measured in both the electron and muon decay channels. Differential cross sections for many observables are also presented including measurements of the jet observables such as the rapidities and the transverse momenta as well as measurements of event observables such as the scalar sums of the transverse momenta of the jets. The measurements are compared to numerous QCD predictions including next-to-leading-order perturbative calculations, resummation calculations and Monte Carlo generators.
}

\newcommand{\TeV}{\ifmmode {\mathrm{\ Te\kern -0.1em V}}\else
                   \textrm{Te\kern -0.1em V}\fi}%

\newcommand{\GeV}{\ifmmode {\mathrm{\ Ge\kern -0.1em V}}\else
                   \textrm{Ge\kern -0.1em V}\fi}%

\newcommand{\MeV}{\ifmmode {\mathrm{\ Me\kern -0.1em V}}\else
                   \textrm{Me\kern -0.1em V}\fi}%

\newcommand{\met}{\ensuremath{E_{\mathrm{T}}^{\mathrm{miss}}}}
\newcommand{\alp}{ALPGEN}
\newcommand{\her}{HERWIG}
\newcommand{\pyt}{PYTHIA}
\newcommand{\she}{SHERPA}
\newcommand{\bh}{{\sc BlackHat}}
\newcommand{\bhs}{{\sc BlackHat}+SHERPA\xspace}

\newcommand{\meps}{MEPS@NLO}

\newcommand{\ace}{AcerMC}
\newcommand{\pow}{POWHEG}
\newcommand{\per}{PERUGIA2011C}
\newcommand{\ctt}{CT10}
\newcommand{\cts}{CTEQ6L1}

\newcommand{\gea}{GEANT4}

\newcommand{\njets}{\ensuremath{{N}_{\mathrm{jets}}}\xspace}

\newcommand{\ipb}{\ensuremath{\rm pb^{-1}}}
\newcommand{\ifb}{\ensuremath{\rm fb^{-1}}}
\newcommand{\mt}{\ensuremath{m_{\mathrm{T}}}}

\newcommand{\pt}{\ensuremath{p_{\rm T}}}

\newcommand{\dr}{\ensuremath{\Delta R}}

\newcommand{\tev}{\ifmmode {\mathrm{\ Te\kern -0.1em V}}\else
                   \textrm{Te\kern -0.1em V}\fi}%
\newcommand{\gev}{\ifmmode {\mathrm{\ Ge\kern -0.1em V}}\else
                   \textrm{Ge\kern -0.1em V}\fi}%
\newcommand{\el}{\ensuremath{e}}
\newcommand{\mgeq}{\ensuremath{\geq}}

\newcommand{\htj}{\ensuremath{H_{\rm T}}}
\newcommand{\stj}{\ensuremath{S_{\rm T}}}

\newcommand{\nj}{\ensuremath{N_{\rm jet}}}
\newcommand{\njs}{\ensuremath{N_{\rm jets}}}
\newcommand{\drjj}{\ensuremath{\Delta R_{j1,j2}}}

\newcommand{\dpjj}{\ensuremath{\Delta\phi_{j1,j2}}}
\newcommand{\dyjj}{\ensuremath{\Delta y_{j1,j2}}}

\newcommand{\Zmm}{\ensuremath{Z\to\mu\mu}}

\newcommand{\Zee}{\ensuremath{Z\to\el\el}}
\newcommand{\Ztt}{\ensuremath{Z\to\tau\tau}}

\newcommand{\Zjets}{\ensuremath{\Zg\,\texttt{+}\,\mathrm{jets}}}

\newcommand{\Wjets}{\ensuremath{W\,\texttt{+}\,\mathrm{jets}}}
\newcommand{\antibar}[1]{\ensuremath{#1\bar{#1}}}
\newcommand{\ttbar}{\antibar{t}\xspace}

\newcommand{\Wg}{\ensuremath{W}}
\newcommand{\Wln}{\ensuremath{W\to\ell\nu}}

\newcommand{\Wen}{\ensuremath{W\to\el\nu}}
\newcommand{\Wmn}{\ensuremath{W\to\mu\nu}}
\newcommand{\Wtn}{\ensuremath{W\to\tau\nu}}
\newcommand{\Zg}{\ensuremath{Z}}

\newcommand{\Wnjh}{\ensuremath{\Wg\,\texttt{+}\mgeq n\texttt{-}\mathrm{jets}}}

\newcommand{\Wno}{\ensuremath{\Wg\,\texttt{+}\mgeq n\, \texttt{+}\, 1}}
\newcommand{\Wj}{\ensuremath{\Wg\,\texttt{+}\mgeq\mathrm{1~jet}}}
\newcommand{\Wjh}{\ensuremath{\Wg\,\texttt{+}\mgeq\mathrm{1}\texttt{-}\mathrm{jet}}}

\newcommand{\Wjeh}{\ensuremath{\Wg\,\texttt{+}\, \mathrm{1}\texttt{-}\mathrm{jet}}}
\newcommand{\Wjj}{\ensuremath{\Wg\,\texttt{+}\mgeq\mathrm{2~jets}}}
\newcommand{\Wjjh}{\ensuremath{\Wg\,\texttt{+}\mgeq\mathrm{2}\texttt{-}\mathrm{jets}}}

\newcommand{\Wjjeh}{\ensuremath{\Wg\,\texttt{+}\, \mathrm{2}\texttt{-}\mathrm{jets}}}

\newcommand{\Wjjj}{\ensuremath{\Wg\,\texttt{+}\mgeq\mathrm{3~jets}}}
\newcommand{\Wjjjj}{\ensuremath{\Wg\,\texttt{+}\mgeq\mathrm{4~jets}}}
\newcommand{\Wjjjjj}{\ensuremath{\Wg\,\texttt{+}\mgeq\mathrm{5~jets}}}

\newcommand{\antikt}{anti-$k_{t}$}

\newcommand{\tight}{``tight''\xspace}

\newcommand{\dressed}{``dressed''\xspace}

\PreprintCoverPaperTitle{Measurements of the \Wg\ production cross sections in association with jets with the ATLAS detector}  % paper title

\PreprintIdNumber{CERN-PH-EP-2014-199}  % CERN preprint number ---> can be found in CDS entry for Draft 2!

\PreprintCoverAbstract{\wjetsabstract}

\PreprintJournalName{Eur.~Phys.~J.~C}

\journalname{Eur. Phys. J. C}
\begin{document}

%\begin{linenumbers}

\title{Measurements of the \Wg\ production cross sections in association with jets with the ATLAS detector}

\author{The ATLAS Collaboration} % \thanksref{e1,addr1}}

%\thankstext{e1}{e-mail: atlas.publications@cern.ch}

\institute{CERN, 1211 Geneva 23, Switzerland \label{addr1}}
\date{Received: date / Accepted: date}

\maketitle

\begin{abstract}
\wjetsabstract
\end{abstract}

%%%%%%%%%%%%%%%%%%%%%%%%%%%%%%%%%%%%%%%%%%%%%%%%%%%%%%%%%%%%%

\section{Introduction}

With the large data sample accumulated in 2011 at the Large Hadron Collider (LHC), detailed investigations of perturbative quantum chromodynamics (pQCD) and electroweak (EWK) effects are now possible over five orders of magnitude in the \Wjets\ production cross section as a function of jet multiplicity and six orders of magnitude as a function of the jet transverse momenta. For the production of a massive gauge boson accompanied by jets, jet transverse momenta up to 1 \TeV\ are now, for the first time, accessible; this is a kinematic region where higher-order EWK effects can become as important as those from higher-order pQCD corrections. During the last few years, advances in the theoretical frameworks for the calculation of final states containing a vector boson and jets allow cross sections to be determined at next-to-leading order (NLO) in pQCD for vector bosons with up to five jets in the final state~\cite{Bern:2013gka}. However, although calculations of EWK effects exist~\cite{Denner:2009gj}, they are not yet incorporated into the theoretical predictions of \Wjets\ production. 

Measurements of \Wjets\ production in proton--anti-proton collisions with a centre-of-mass energy of $\sqrt{s}=1.96 \TeV$ have been reported by the CDF and D0 collaborations~\cite{Aaltonen:2007ip,Abazov:2013gpa} and for $\sqrt{s}=7\TeV$ proton--proton collisions using an integrated luminosity of 35 \ipb\ by the ATLAS collaboration~\cite{Aad:2012en} and 5.0 \ifb\ by the CMS collaboration~\cite{Khachatryan:2014uva}. This paper presents updated and extended measurements of \Wjets\ production in proton--proton collisions at $\sqrt{s}=7 \TeV$ by the ATLAS collaboration using an integrated luminosity of 4.6 \ifb\ collected in 2011 and includes detailed comparisons to a number of new theoretical predictions. The results in this paper are based on both the \Wen\ and \Wmn\ decay channels. 

The paper is organised as follows. The ATLAS detector is described in Sect.~\ref{sec:Detector}. Section~\ref{sec:simulation} provides details of the simulations used in the measurement.  A description of the data set, the electron and muon selection, the selection of \Wjets\ events and the background estimation is given in Sect.~\ref{sec:data-sel}. The procedure used to correct the measurements for detector effects and the combination of the electron and muon results are described in Sect.~\ref{sec:Unfolding}. The treatment of the systematic uncertainties is detailed in Sect.~\ref{sec:sys}. Section~\ref{sec:Theory} provides a description of the NLO pQCD predictions and corrections applied to them. Section~\ref{sec:results} discusses the results. Finally Sect.~\ref{sec:summary} provides conclusions.

\section{ATLAS detector}
\label{sec:Detector}

The ATLAS detector~\cite{Aad:2008zzm} is a multi-purpose detector with a
symmetric cylindrical geometry and nearly $4\pi$
coverage in solid angle.\footnote{ATLAS uses a right-handed coordinate
  system with its origin at the nominal interaction point (IP) in the
  centre of the detector and the $z$-axis along the beam pipe.  The
  $x$-axis points from the IP to the centre of the LHC ring, and the
  $y$-axis points upward.  Cylindrical coordinates ($r$, $\phi$) are
  used in the transverse plane, $\phi$ being the azimuthal angle
  around the beam pipe.  The pseudorapidity is defined in terms of theŒ
  polar angle $\theta$ as $\eta = -\ln \tan(\theta$/2).}
The collision point is surrounded by inner tracking devices, which in increasing radii are followed by a
superconducting solenoid providing a 2~T magnetic field, a calorimeter
system, and a muon spectrometer.  In order of increasing radii, the inner tracker consists of silicon pixel and microstrip detectors and a transition radiation tracker, and provides precision tracking for charged particles in the pseudorapidity range $|\eta| < 2.5$.  
 The calorimeter system has liquid argon (LAr) or
scintillator tiles as the active media.  In the pseudorapidity region
$|\eta| < 3.2$, high-granularity LAr electromagnetic (EM) sampling
calorimeters are used.  A scintillator tile calorimeter provides
hadronic coverage for $|\eta| < 1.7$.  The endcap and forward regions,
spanning $1.5<|\eta| <4.9$, are instrumented with LAr calorimeters for
both the EM and hadronic measurements.  The muon spectrometer consists of
three large superconducting toroids each consisting of eight coils and a system
of trigger chambers and precision tracking chambers which provide
triggering and tracking capabilities in the ranges $|\eta| < 2.4$ and
$|\eta| < 2.7$, respectively. A three-level trigger system is used to select interesting events~\cite{Aad:2012xs}. 
The Level-1 trigger reduces the event rate to less than 75 kHz using hardware-based trigger algorithms acting on a subset of detector information. 
Two software-based trigger levels further reduce the event rate to about 400 Hz using the complete detector information.

\section{Simulated event samples}
\label{sec:simulation}

Simulated event samples are used for some of the background estimates, for the correction of the signal yield for detector effects and for comparisons of the results to theoretical expectations. 

Samples of $W \rightarrow \ell \nu$ and $Z \rightarrow \ell \ell$ $(\ell = e,\, \mu,\, \tau)$ events with associated jets are generated with both \alp\ v2.13~\cite{Mangano:2002ea} and {\she} v1.4.1~\cite{Gleisberg:2008ta,Hoeche:2012yf}.
For the \alp\ samples, the matrix element implemented in this generator produces events with up to five additional partons in the final state and is 
interfaced to \her\ v6.520~\cite{Corcella:2000bw,Marchesini:1991ch} for
parton showering and fragmentation, with {\sc JIMMY} v4.31~\cite{Butterworth:1996zw} for underlying event contributions and with 
PHOTOS~\cite{Golonka:2005pn} to calculate final-state radiation from quantum electrodynamics (QED). \alp\ uses the MLM matching scheme~\cite{Mangano:2002ea} to remove any double counting between the matrix element and parton shower calculations.
The \mbox{\cts~\cite{Pumplin:2002vw}} parton distribution functions (PDFs) are
used with the \mbox{AUET2-CTEQ6L1} set of generator parameters (tune)~\cite{ATLAS:2011gmi}. \alp\ samples including heavy-flavour production, such as $W+b\bar{b}$, $W+c\bar{c}$ and $W+c$ production, are used in the estimate of the \ttbar\ background. 
Samples of $W \rightarrow \ell \nu$ are also produced with \alp\ v2.14 interfaced to \pyt\ v6.425~\cite{Sjostrand:2006za}
using the \per~\cite{Skands:2010ak}~tune and are used to estimate the uncertainties due to non-perturbative effects, as described in Sect.~\ref{sec:nonpert}.
Samples of $W \rightarrow \ell \nu$ are also produced using {\she}, which uses the CKKW~\cite{Catani:2001cc} matching scheme, \ctt\ PDFs~\cite{Lai:2010vv} and an internal model for QED radiation based on the YFS method~\cite{Yennie:1961ad}. These samples are generated with up to four additional partons.  

Top quark pair production is simulated with \alp\ interfaced to \her, using the same configuration as for the \Wg\ samples. Additional  \ttbar\ samples are generated with the \pow-Box v1.0 generator~\cite{Alioli:2010xd}, interfaced to \pyt\ using the \per\ tune and configured to use CT10 PDFs. Single top quark production, including $Wt$ production, is modelled with {\ace} 3.8~\cite{Kersevan:2004yg} with 
MRST LO* PDFs~\cite{Sherstnev:2007nd}, interfaced to \pyt.  
The diboson production processes $WW, WZ$, and $ZZ$ are generated with {\her} v6.510, interfaced to {\sc JIMMY} v4.3 and using MRST LO* PDFs and the {\sc AUET2-LO*} tune~\cite{ATLAS:2011gmi}. 

The generated samples are passed through a simulation of the ATLAS detector based on \gea~\cite{Agostinelli:2002hh,Aad:2010ah} and through
a trigger simulation.  The simulated samples are overlaid with additional proton--proton interactions (``pile-up'') generated
with \pyt\ using the AMBT1 tune~\cite{Aad:2010ac} and the distribution of the average number of interactions per bunch crossing is
reweighted to agree with the corresponding data distribution. The simulated events are reconstructed and analysed with the same analysis chain as for the data. Scale factors are applied to the simulated samples to correct for the small differences from data in the  trigger, reconstruction and identification efficiencies for electrons and muons.

All samples are normalised to the respective inclusive cross sections calculated at higher orders in pQCD.  The \Wg\ and \Zg\ samples are normalised to the
next-to-next-to-leading-order (NNLO) pQCD inclusive predictions calculated with the FEWZ~\cite{Melnikov:2006kv} program and MSTW2008
NNLO PDFs~\cite{Martin:2009iq}.  The \ttbar\ cross section is calculated at NNLO+NNLL as in Refs.~\cite{Baernreuther:2012ws, Czakon:2012zr, Czakon:2012pz, Cacciari:2011hy, Czakon:2013goa, Czakon:2011xx} and the diboson cross sections are calculated at NLO using {\sc MCFM}~\cite{mcfm_dib} with MSTW2008 PDFs.

\section{Data selection and event analysis}
\label{sec:data-sel}

The data used in this analysis were collected during the 2011 LHC proton--proton collision run at a centre-of-mass energy of $\sqrt{s}= 7 \TeV$.  
After application of beam and data-quality requirements, the total integrated luminosity is 4.6 \ifb\ with an uncertainty of 1.8\%~\cite{Aad:2013ucp}. 

Events are selected for analysis by requiring either a single-electron or single-muon trigger. The single-electron trigger required an electron with a transverse momentum (\pt) greater than $20 \GeV$ for the first 1.5 \ifb\ of data and a transverse momentum greater than 22~\GeV\ for the remaining 3.1~\ifb\ of data. The single-muon trigger required a muon with a transverse momentum greater than 18~\GeV. For both the electron and muon triggers, the thresholds are low enough to ensure that leptons with $\pt > 25 \GeV$ lie on the trigger efficiency plateau. 

In both decay channels, events are required to have at least one reconstructed vertex with at least three associated tracks, where the tracks must have a \pt\ greater than 400~\MeV. The vertex with the largest $\Sigma \pt^2$ of associated tracks is taken as the primary vertex. 

\subsection{Electron reconstruction and identification}

Electrons are reconstructed from energy clusters in the calorimeter and matched to an inner detector track. They are required to satisfy a set of identification criteria. This so-called \tight selection is similar to the one defined in Ref.~\cite{Aad:2014fxa}. The \tight selection includes requirements on the transverse impact parameter with respect to the primary vertex and on the number of hits in the innermost pixel layer in order to reject photon conversions. The electron must have \pt~$> 25 \GeV$ and $|\eta| < 2.47$ and electrons in the transition region between the barrel and endcap calorimeter ($1.37<|\eta|<1.52$) are rejected.  Events are rejected if there is a second electron passing the same selection as above. In order to suppress background from events where a jet is misidentified as an electron, the electron is required to be isolated. A \pt- and $\eta$-dependent requirement on a combination of calorimeter and track isolation variables is applied to the electron, in order to yield a constant efficiency across different momentum ranges and detector regions. The track-based isolation uses a cone size of $\dr \equiv \sqrt{ (\Delta\phi)^{2} + (\Delta \eta)^{2} } = 0.4$ and the calorimeter-based isolation uses a cone size of $\dr = 0.2$. The actual requirements on the maximum energy or momentum allowed in the isolation cone range between 2.5~\GeV\ and 4.5~\GeV\ for the calorimeter-based isolation and between 2.0~\GeV\ and 3.0~\GeV\ for the track-based isolation.

\subsection{Muon reconstruction and identification}

Muons are required to be reconstructed by both the inner detector and muon spectrometer systems~\cite{Aad:2014zya} and to have \pt~$>$ 25~\GeV\ and $|\eta| <2.4$. 
Events are rejected if there is a second muon passing the same kinematic selections as above. 
As in the electron channel, an isolation criterion is applied to reduce the background of semileptonic heavy-flavour decays. The track-based isolation fraction, which is defined as the summed scalar \pt\ of all tracks within a cone size of $\dr = 0.2$ around the muon, divided by the \pt\ of the muon itself, $\Sigma \pt^{\rm{tracks}}/ \pt^{\rm {muon}}$, must be less than 10\%. To further reject events from semileptonic heavy-flavour decays, the transverse impact parameter significance of the muon with respect to the primary vertex is required to satisfy $|d_0/\sigma(d_0)| < 3.0$ where $d_0$ is the muon impact parameter and $\sigma(d_0)$ is the estimated per-track uncertainty on $d_0$.  

\newcommand{\plotsetDetTwo}[3]{
\begin{figure*} \centering
\includegraphics[width=0.45\textwidth]{plot_#1_we_log} 
\includegraphics[width=0.45\textwidth]{plot_#1_wm_log} 
\caption{Distribution of events passing the \Wjets\ selection as a function of the #2 for the electron (left) and muon (right) channels. On the data points, the statistical uncertainties are smaller than the size of the points and the systematic uncertainties, described in Sect.~\ref{sec:sys}, are shown by the hashed bands whenever visible. The lower panel shows ratios of the predictions for signal and background to the data, where either \alp\ (black line) or \she\ (red dashed line) is used for the signal simulation. The experimental systematic uncertainties are shown by the yellow (inner) band and the combined statistical and systematic uncertainties are shown by the green (outer) band.}
\label{#3}
\end{figure*}
}

\subsection{Jet selection}
	
Jets are reconstructed using the \antikt~algorithm \cite{Cacciari:2008gp} with a radius parameter $R = 0.4$ using topological clusters~\cite{Aad:2014bia} of energy depositions in the calorimeters as input.  
Jets arising from detector noise or non-collision events are rejected. 
To take into account the differences in calorimeter response to electrons and hadrons and to correct for inactive material and out-of-cone effects, \pt- and $\eta$-dependent factors, derived from a combination of simulated events and in situ methods~\cite{Aad:2014bia}, are applied to each jet to provide an average energy scale correction. The jet energies are also corrected to account for energy arising from pile-up. 

Jets are required to have $\pt> 30 \GeV$ and a rapidity of $|y| < 4.4$. Rapidity is defined as $\frac{1}{2} \ln [ (E+p_z)/(E-p_z)]$, where $E$ denotes the energy and $p_z$ is the component of the momentum along the beam direction.  All jets within $\dr= 0.5$ of an electron or muon that passed the lepton identification requirements are removed. In order to reject jets from additional proton-proton interactions, the summed scalar \pt\ of tracks which are associated with the jet and associated with the primary vertex is required to be greater than  75\% of the summed \pt\  of all tracks associated with the jet. This criterion is applied to jets within the acceptance of the tracking detectors, $|\eta|<2.4$.  
The residual impact of pile-up on the distribution of the jet observables was studied by comparing data and simulation for different data periods. The simulation was found to reproduce well the pile-up conditions.

\subsection{\Wg\ selection}

For both the \Wen\ and \Wmn\ selections, events are required to have a significant missing transverse momentum (\met) and large transverse mass (\mt).  The latter is defined by the lepton and neutrino \pt\ and direction as $\mt = \sqrt{2 \pt^{\ell} \pt^{\nu}(1-\cos(\phi^{\ell}-\phi^{\nu}))}$,
where the $(x,y)$ components of the neutrino momentum are those of the missing transverse momentum. The \met\ is calculated as the negative vector sum of the transverse momenta of calibrated leptons, photons and jets and additional low-energy deposits in the calorimeter~\cite{Aad:2012re}. Events are required to have \met~$>$ 25 GeV and $\mt > 40$~\GeV.

\subsection{Background}

In both the electron and muon channels, the background processes include \Wtn~where the $\tau$ decays to an electron or muon, \Zee\ or \Zmm\ where one lepton is not identified, \Ztt, leptonic \ttbar\ decays ($\ttbar \rightarrow b\overline{b} q q' \ell \nu$ and $\ttbar \rightarrow b\overline{b} \ell \nu \ell \nu$), single-top, diboson ($WW$, $WZ$, $ZZ$) and multijet events. The multijet background in the electron channel has two components: one where a light-flavour jet passes the electron selection and additional energy mismeasurement in the event results in large \met\ and another where an electron is produced from a  semileptonic decay of a bottom- or charm-hadron. For the muon channel, the multijet background arises from semileptonic heavy-flavour decays. 
 
 At small numbers of associated jets ($\njs$), the dominant background arises from multijet events while at high multiplicities \ttbar\ events are dominant. Using the event selection defined above, the multijet background constitutes 11\% of $\njs=1$ events and the \ttbar\ background is 80\% of $\njs=7$ events. The \ttbar\ background can be reduced by applying a veto on events with $b$-jets. However, the selection in this analysis was kept as inclusive as possible to allow for direct comparison with measurements of \Zjets\ production~\cite{Aad:2013ysa}, to be used in the determination of the ratio of \Wjets\ to \Zjets\ production~\cite{atlas-rjet}, and to minimise theoretical uncertainties in the fiducial cross-section definition. For the multijet and \ttbar\ background, data-driven methods are used to determine both the total number of background events in the signal region as well as the shape of the background for each of the differential distributions. 

The number of multijet background events is estimated by fitting, in each jet multiplicity bin, the \met~distribution in the data (with all selection cuts applied except the cut on \met)  to a sum of two templates: one for the multijet background and another which includes the signal and other background contributions. In both the muon and electron channels, the shape for the first template is obtained from data while the second template is from simulation. 
To select a data sample enriched in multijet events in the electron channel, dedicated electron triggers with loose identification criteria and
additional triggers requiring electrons as well as jets are used.
The multijet template is built from events which fail the \tight requirements 
of the nominal electron selection in order to suppress signal contamination. 
Electrons are also required to be
non-isolated in the calorimeter, i.e. they are required to have an energy
deposition in the calorimeter in a cone of $\dr = 0.3$ centred on the electron direction 
larger than $20 \%$ of the total transverse energy of the electron.
In the muon channel, the multijet template is also obtained from data, by selecting events where the scalar sum \pt\ of all tracks within a cone of size $\dr = 0.2$ around the muon is between $10\%$ and $50\%$ of the muon \pt. 

In both channels, the sample used to extract the template for the multijet background is statistically independent of the signal sample. 
The fit is performed for each jet multiplicity up to five-jet events. Due to fewer events in the multijet template for six- and seven-jet events, the number of multijet events is determined by performing a single fit for events with five or more jets.

At high multiplicities, the background from \ttbar~events is larger than the signal itself. Although \ttbar\ simulations can be used to estimate this background, a data-driven approach is used in order to reduce the systematic uncertainties. Using a similar method to that used for the multijet background determination, the number of \ttbar\ events is estimated by fitting a discriminant distribution in the data to the sum of three templates: the \ttbar\ template, the multijet template and one which includes the signal and remaining background contributions. The discriminant variable chosen is the transformed aplanarity, defined as e$^{(-8\, A)}$, where $A$, the aplanarity, is 1.5 times the smallest eigenvalue of the normalised momentum tensor as defined in Ref.~\cite{Aad:2012qf}. By definition, an isotropic event has an aplanarity of one half, whereas a planar event has a value of zero. Since \ttbar\ events are more isotropic than the \Wjets\ signal, the transformed aplanarity was found to yield good separation between the signal and background with small systematic uncertainties on the background estimate. For the aplanarity calculation, the lepton and all jets passing the selection are used in the momentum tensor. The multijet template is as described above and the \Wg\ signal template is taken from simulations. The \ttbar~template is derived from a control region in data by requiring at least one $b$-tagged jet in the event. A multivariate $b$-tagging algorithm was used at a working point with a 70\% $b$-tagging efficiency~\cite{mv1}. With this selection, the \ttbar\ control region has a purity of 60\% in events with three jets and 97\% in events with six jets. Non-\ttbar~events passing the selection, such as \Wg+light-jets, $W+b$, $W+c$ and $b$-tagged multijet events are subtracted from the \ttbar\ control region using simulations or in the case of the multijet events using the fit to \met\ as described above but with an event sample where the $b$-tagging requirement has been applied. Since $b$-tagging is only available for jets within $|y| < 2.4$ where information from the tracking detectors exists, the $b$-tagging selection biases some of the kinematic distributions, most notably the jet rapidity distribution. To account for this, \ttbar~simulations are used to correct for any residual bias. The corrections are a few percent in most regions but up to 30\% at very high jet rapidities. The fits to the transformed aplanarity distribution are performed for each exclusive jet multiplicity from three to six jets. 
In the fit, the normalisation of the multijet background is obtained from the \met\ fit above. The estimated number of \ttbar\ events is consistent with the predictions from \ttbar\ simulations for all distributions and the uncertainties from the data-driven method are smaller than those from the simulations. Since the \ttbar~template is a sub-sample of the signal data sample, there is a statistical correlation to the signal sample.  This is estimated using pseudo datasets derived via Poisson variations of the signal and \ttbar\ simulated samples and is found to be 15\% at \njs$=3$ and 45\% at \njs$=6$. The fit uncertainties are corrected to account for this correlation. For lower multiplicities of $\njs \le2$, where the fraction of \ttbar\ is less than 5\%, simulations are used for the background estimate. 

The remaining background contributions are estimated with simulated event samples. These background samples are normalised to the integrated luminosity of the data using the cross sections as detailed in Sect.~\ref{sec:simulation}.

\subsection{Reconstruction-level results}

The measured and expected distributions of the jet observables are compared at the reconstruction level, separately in the electron and muon channels, using the selection criteria described above. Some example distributions, namely the inclusive jet multiplicity, the \pt\ and rapidity of the highest-\pt\ (leading) jet and the summed scalar  \pt\ of the lepton and all jets 
plus \met\ (called \htj) are shown in Figs.~\ref{fig:det-njet}--\ref{fig:det-ht}. The data are consistent with the predictions from the \alp\ and \she\ generators. The numbers of selected events including the estimated background contributions are summarised in Table~\ref{tab:event-sel} for both the electron and muon channels. 

\begin{table*}[hbt] 
\renewcommand{\arraystretch}{1.0}
\centering
    \begin{tabular}{%
c|
rrrrrrrr
%S[table-number-alignment=right]
%S[table-number-alignment=right]
%S[table-number-alignment=right]
%S[table-number-alignment=right]
%S[table-number-alignment=right]
%S[table-number-alignment=right]
    }
    \toprule
$N_{\mbox{jet}} $ & 0  & 1 &  2  &  3  & 4 & 5 & 6 & 7 \\
    \midrule    
 &  \multicolumn{7}{c}{ $W \rightarrow e\nu$  } \\ 
    \midrule
$W\rightarrow e\nu$ &	94\%	&	78\%	&	73\%	&	58\%	&	37\%	&	23\%	&	14\%	&	11\%	\\
Multijet & 	4\%	&	11\%	&	12\%	&	11\%	&	7\%	&	6\%	&	5\%	&	4\%	\\
\ttbar  &	$<1$\%	&	$<1$\%	&	3\%	&	18\%	&	46\%	&	62\%	&	76\%	&	80\%	\\
Single top  &	$<1$\%	&	$<1$\%	&	2\%	&	3\%	&	4\%	&	3\%	&	2\%	&	2\%	\\
$W\rightarrow \tau \nu$, diboson &	2\%	&	3\%	&	3\%	&	3\%	&	2\%	&	1\%	&	1\%	&	1\%	\\
$Z\rightarrow ee$  &	$<1$\%	&	8\%	&	7\%	&	7\%	&	5\%	&	4\%	&	3\%	&	3\%	\\
\midrule
  Total Predicted  & \ytot{11089032} & \ytot{1511675} & \ytot{353942} & \ytot{89527} & \ytot{28177} & \ytot{8545} & \ytot{2534} & \ytot{572} \\
                            &  \yerr{ 642724} & \yerr{ 98769} & \yerr{ 23450} & \yerr{ 5632} & \yerr{ 1380} & \yerr{ 435} & \yerr{ 199} & \yerr{ 61} \\
    \midrule    
 Data Observed  & \num{ 10878398 } & \num{ 1548000 } & \num{ 361957 } & \num{ 91212 } & \num{ 28076 } & \num{ 8514 } & \num{ 2358 }  & \num{618}\\

    \midrule    
  & \multicolumn{7}{c}{ $W \rightarrow \mu\nu$ } \\ 
    \midrule    
$W\rightarrow \mu\nu$ &	93\%	&	82\%	&	78\%	&	62\%	&	40\%	&	25\%	&	17\%	&	11\%	\\
Multijet &	2\%	&	11\%	&	10\%	&	9\%	&	7\%	&	5\%	&	4\%	&	3\%	\\
\ttbar  &	$<1$\%	&	$<1$\%	&	3\%	&	19\%	&	46\%	&	64\%	&	75\%	&	83\%	\\
Single top &	$<1$\%	&	$<1$\%	&	2\%	&	3\%	&	4\%	&	3\%	&	2\%	&	2\%	\\
$W\rightarrow \tau \nu$, diboson &	2\%	&	3\%	&	3\%	&	3\%	&	2\%	&	1\%	&	1\%	&	$<1$\%	\\
$Z\rightarrow \mu\mu$ &	3\%	&	4\%	&	3\%	&	3\%	&	2\%	&	1\%	&	1\%	&	1\%	\\
\midrule
  Total Predicted          & \ytot{13344999} & \ytot{1705186}    & \ytot{384352} & \ytot{96702}  & \ytot{30075} & \ytot{8985} & \ytot{2397} & \ytot{627} \\ 
                                     & \yerr{770189}      & \yerr{104353}         & \yerr{23864}  & \yerr{6128} & \yerr{1640} & \yerr{483} & \yerr{181} & \yerr{66}\\ 
\midrule 
 Data Observed & \num{13414400} & \num{1758239} & \num{403146} & \num{99749} & \num{30400} & \num{9325} & \num{2637} & \num{663} \\
\bottomrule 
\end{tabular}
\caption{The approximate size of the signal and backgrounds, expressed as a fraction of the total number of predicted events. They are derived from either data-driven estimates or simulations for exclusive jet multiplicities for the $W \rightarrow e \nu$ selection (upper table) and for the $W \rightarrow \mu\nu$ selection (lower table). The total numbers of predicted and observed events are also shown. }
\label{tab:event-sel}

 \end{table*}
 
 \plotsetDetTwo{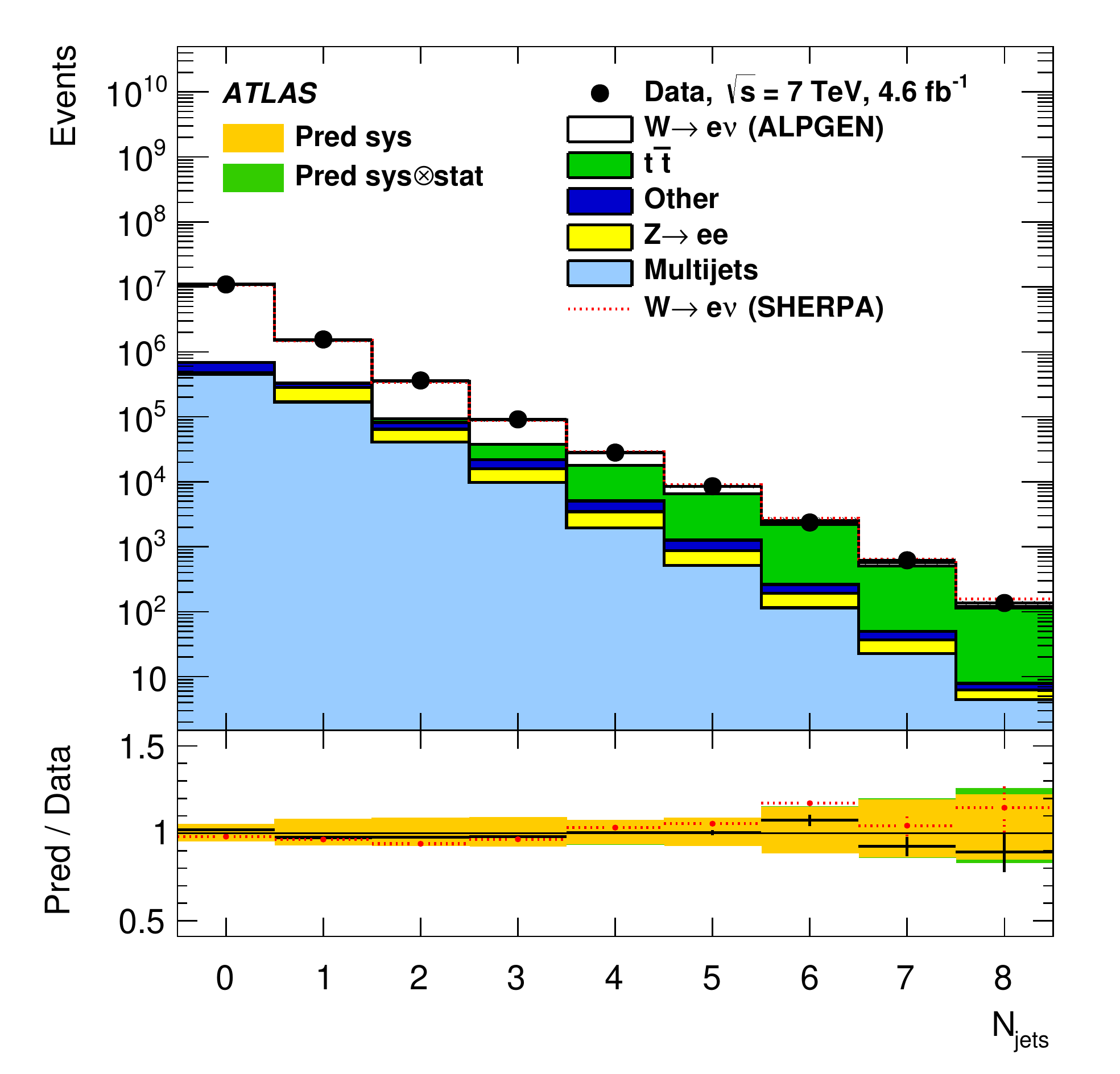}{inclusive jet multiplicity (\njs)}{fig:det-njet}
 \plotsetDetTwo{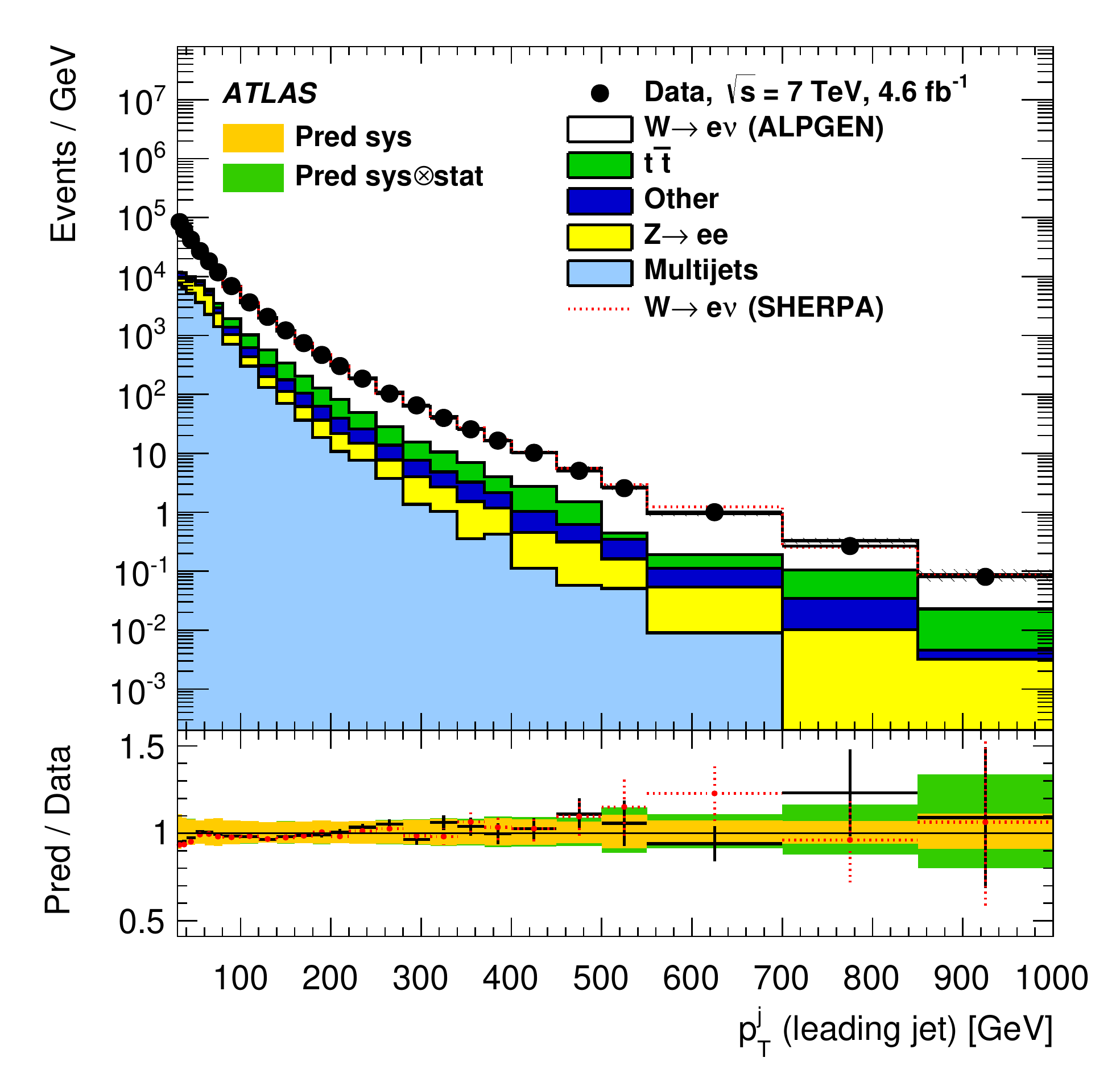}{leading jet \pt~}{fig:det-pt}
 \plotsetDetTwo{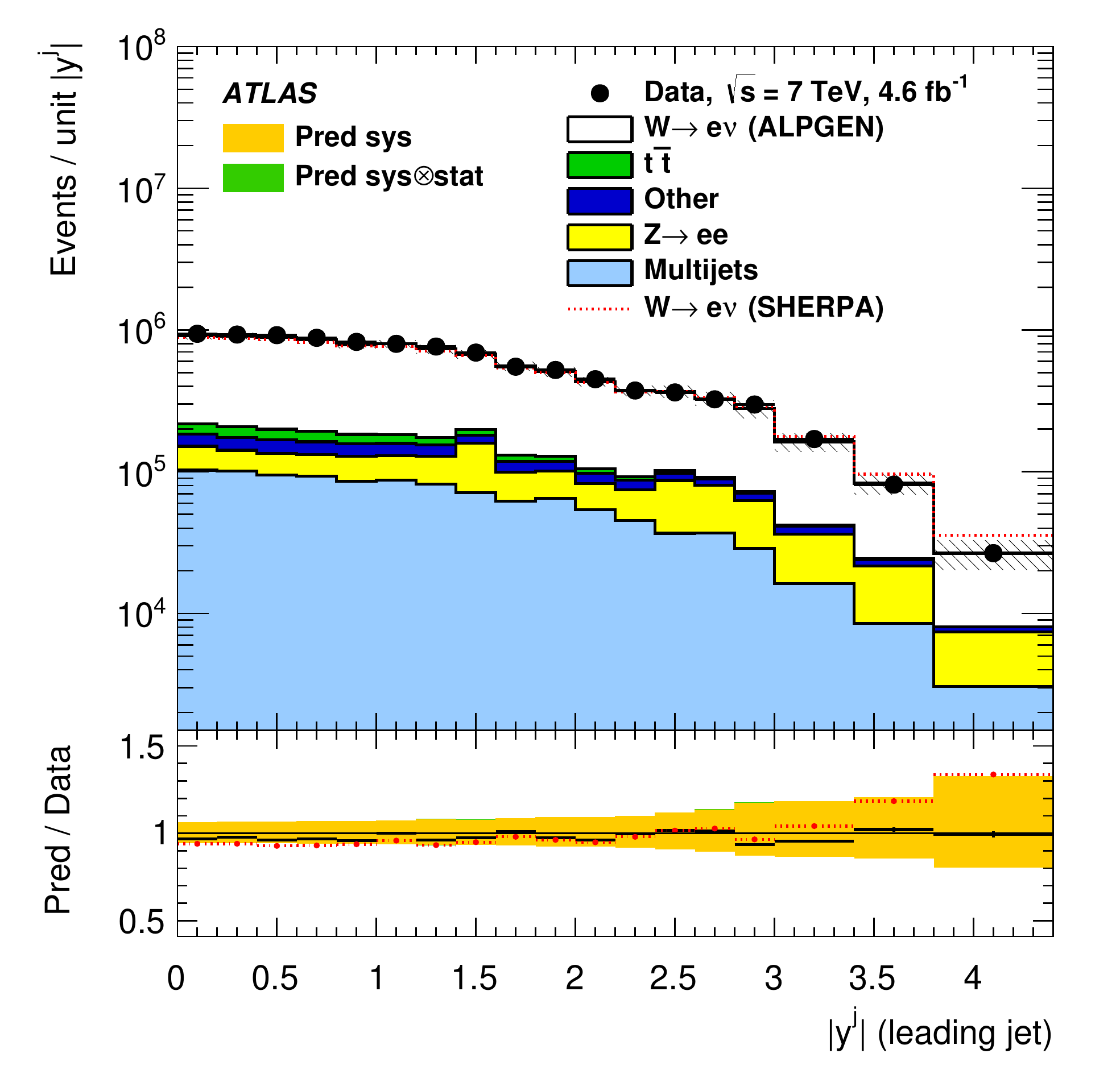}{leading jet rapidity}{fig:det-rap}
 \plotsetDetTwo{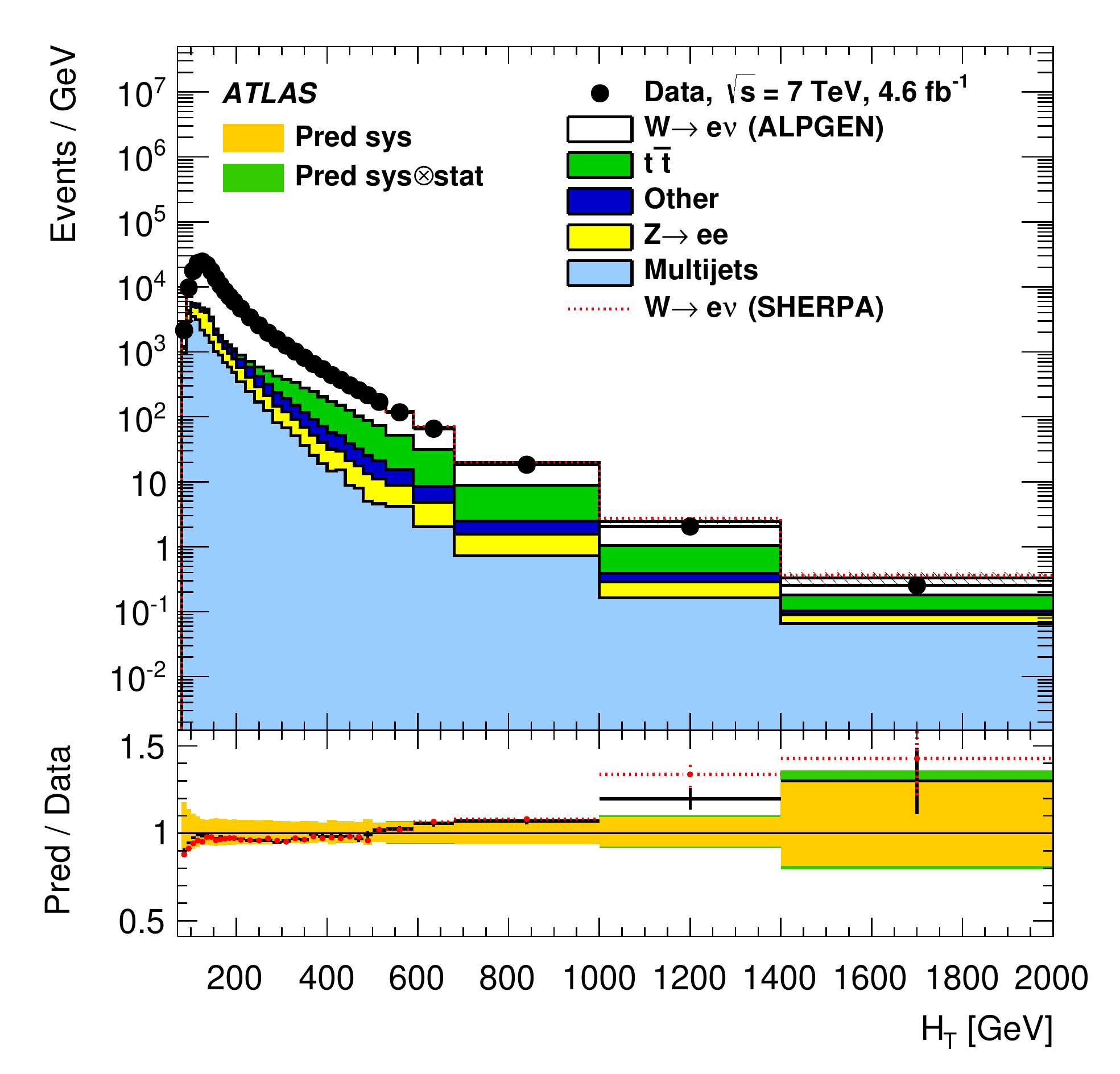}{summed scalar \pt\ of all identified objects in the final state, \htj}{fig:det-ht}

\begin{table*}[htb]
\begin{center}
\begin{tabular}{lccc} 
%\hline 
\toprule 
  & Electron Channel & Muon Channel & Combined \\ 
\midrule 
Lepton \pt & $\pt > 25$~\GeV & $\pt > 25$~\GeV & $\pt > 25$~\GeV \\ 
Lepton rapidity & $|\eta|<2.47$ (excluding $1.37<|\eta|<1.52$) &$|\eta|<2.4$ & $|\eta|<2.5$ \\
\midrule
&
\multicolumn{3}{c}{\Wln\ criteria} \\ \hline $Z$ veto &
\multicolumn{3}{c}{exactly one lepton} \\ Missing transverse momentum &
\multicolumn{3}{c}{$\met > 25$~\GeV} \\ Transverse mass &
\multicolumn{3}{c}{$\mt > 40$~\GeV} \\ 
\midrule
&
\multicolumn{3}{c}{Jet criteria} \\ 
\midrule 
Jet \pt & \multicolumn{3}{c}{$\pt > 30$~\GeV} \\ 
Jet rapidity & \multicolumn{3}{c}{|$y| < 4.4$} \\ 
Jet isolation & \multicolumn{3}{c}{$\dr(\ell,{\rm jet}) > 0.5$ (jet is removed)} \\ 
\bottomrule
%\hline
\end{tabular}
\caption{Kinematic criteria defining the fiducial phase space at particle level for the \Wen\ and \Wmn\ channels as well as the combination. The \Wln\ and jet criteria are applied to the electron and muon channels as well as the combination. 
\label{tab:selection}}
\end{center}
\end{table*} 

\begin{table*}[htb] 
\centering
\begin{tabular}{l|rrrrrrrr}

\toprule
 \multicolumn{1}{c}{ $(\Wen)$ } & 
 \multicolumn{1}{c}{ Incl. } & 
 \multicolumn{1}{c}{ $\njs \ge 1$ } & 
 \multicolumn{1}{c}{ $\njs \ge 2$ } & 
 \multicolumn{1}{c}{ $\njs \ge 3$ } & 
 \multicolumn{1}{c}{ $\njs \ge 4$ } &
 \multicolumn{1}{c}{ $\njs \ge 5$ } & 
 \multicolumn{1}{c}{ $\njs \ge 6$ } &
 \multicolumn{1}{c}{ $\njs \ge 7$ } \\ 
\hline
Electron &     1.1\% &     1.3\% &     1.3\% &     1.2\% &     1.2\% &     1.3\% &     2.7\% & 3.4\%  \\
Jets &     0.3\% &     9\% &    11\% &    15\% &    20\% &    29\% &    42\% & 45\% \\
\ttbar backgrounds &     $<0.1$\% &     0.2\% &     1.0\% &     4.8\% &    13\% &    39\% &   100\% & 90\%  \\
Multijet backgrounds &     0.5\% &     1.5\% &     2.1\% &     2.1\% &     5\% &    15\% &    25\% & 25\%  \\
\met &   0.2\% &     1.7\% &     1.2\% &     1.2\% &     1.0\% &     0.7\% &     1.7\% & 2.6\%  \\
Unfolding  &     0.2\% &     1.7\% &     0.9\% &     1.1\% &     1.2\% &     0.9\% &     5\% & 22\% \\
Luminosity  &     1.9\% &     2.1\% &     2.1\% &     2.2\% &     2.3\% &     2.5\% &     2.6\% & 2.2\%  \\
\midrule
Total Syst.  &     2.3\% &     10\% &    12\% &    16\% &    25\% &    50\% &   110\%  & 110\% \\
\midrule
\midrule
\multicolumn{1}{c}{ $(\Wmn)$ } & \\ 
\midrule
Muon &     1.5\% &     1.7\% &     1.7\% &     1.4\% &     1.5\% &     2.1\% &     3.7\%  & 4.4\% \\
Jets &    0.1\% &     8\% &     9\% &    13\% &    16\% &    20\% &    29\% & 60\% \\
\ttbar backgrounds &     $<0.1$\% &     0.2\% &     0.9\% &     4.1\% &    11\% &    26\% &    47\% & 60\% \\
Multijet backgrounds &    0.1\% &     0.5\% &     0.8\% &     1.4\% &     2.2\% &     4.2\% &     4.6\%   & 9\% \\
\met  &    0.3\% &     1.0\% &     0.9\% &     1.0\% &     1.0\% &     0.6\% &     0.9\% & 1.1\% \\
Unfolding  &     0.2\% &     1.7\% &     0.9\% &     1.0\% &     1.2\% &     1.3\% &     2.6\%   & 11\% \\
Luminosity &     1.9\% &     2.0\% &     2.0\% &     2.1\% &     2.1\% &     2.1\% &     2.0\%   & 2.0\% \\
\midrule
Total Syst. &     2.5\% &     8\% &     10\% &    14\% &    20\% &    34\% &    60\%   & 80\% \\
\bottomrule
\end{tabular}
\caption{Systematic uncertainties on the measured \Wjets\ cross section in the electron and muon channels as a function of
the inclusive jet multiplicity in percent.}
\label{tab:systematics}
\end{table*}

\section{Corrections for detector effects and combination of channels}
\label{sec:Unfolding}

The yield of signal events is determined by first subtracting the estimated background contributions from the data event counts. 
In each channel the data distributions are then corrected for detector effects to 
the fiducial phase space, defined in Table~\ref{tab:selection}. 
In this definition, the lepton kinematics in the simulation at particle level are based on final-state leptons from the \Wg\ boson decays including the
contributions from the photons radiated by the decay lepton within a cone of $\dr = 0.1$ around its direction (\dressed leptons). 
In the simulation the \met\ is determined from the neutrino from the decay of the \Wg\ boson. Particle-level jets are defined using an \antikt\ algorithm with a radius parameter of $R=0.4$, $\pt > 30 \GeV$ and $|y| < 4.4$. All jets within $\dr = 0.5$ of an electron or muon are removed. Final-state particles with a lifetime longer than 30 ps, either produced directly in the
proton--proton collision or from the decay of particles with shorter lifetimes, are included in the particle-level jet reconstruction. The  neutrino and the electron or  muon from the \Wg\ boson decay, and any photon included in the dressed lepton, are not used for the jet finding. 

The correction procedure is based on samples of simulated events and corrects for
jet and \Wg\ selection efficiencies and resolution effects. 
The correction is implemented using an iterative Bayesian method of
unfolding~\cite{D'Agostini:1994zf}. Simulated events are
used to generate for each distribution a response matrix to account for bin-to-bin migration effects
 between the reconstructed and particle-level distributions.
 The particle-level prediction from simulation is used as an initial prior 
to determine a first estimate of the unfolded data distribution. 
For each further iteration the estimator for the
unfolded distribution from the previous iteration is used as a new input prior.
The bin sizes in each distribution are
chosen to be a few times larger than the resolution of the
corresponding variable. 
The \alp\ \Wjets\ samples provide a
satisfactory description of distributions in data and are employed to
perform the correction procedure.  
The number of iterations was optimised to find a balance between too many iterations, causing high statistical uncertainties associated with the unfolded spectra, and too few iterations, which increases the dependency on the Monte Carlo prior. The optimal number of iterations is typically between one and three, depending on the observable. Since the differences in the unfolded results are negligible over this range of iterations, two iterations were consistently used for unfolding each observable.

The unfolded cross sections measured in the electron and muon channels are then extrapolated to a common lepton phase space region, defined by lepton $\pt > 25 \GeV$ 
and $|\eta | < 2.5$ and summarised in Table~\ref{tab:selection}.  
The extrapolations to the common phase-space are performed using bin-by-bin correction factors, derived from \alp\ \Wjets\ simulated samples described in Sect.~\ref{sec:simulation}. The correction factors are approximately 1.08 and 1.04 for the electron and muon channel cross sections respectively. The extrapolated cross sections measured in the electron and muon channels are in agreement for all observables considered.

The measured differential \Wjets\ production cross sections in the electron and muon channels are combined by averaging using a statistical procedure~\cite{Glazov:2005rn,Aaron:2009bp} that accounts for correlations between the sources of systematic uncertainty affecting each channel. Correlations between bins for a given channel are also accounted for. Each distribution is combined separately by minimising a $\chi^2$ function. 

The combination of the systematic uncertainties for the two channels is done in the following way. The uncertainties on the modelling in the unfolding procedure, the luminosity, all the background contributions estimated from simulations (except for the \Zjets\ background as discussed below) and systematic uncertainties on the data-driven \ttbar\ estimation have been treated as correlated among bins and between channels. The lepton systematic uncertainties are assumed to be correlated between bins of a given distribution, but independent between the two lepton channel measurements. The statistical uncertainties of the data, the statistical uncertainty from the simulations used in the unfolding procedure, and the statistical uncertainty from the \ttbar\ fit are treated as uncorrelated among bins and channels. The systematic uncertainties on the multijet background, which contains correlated and uncorrelated components, are also treated as uncorrelated among bins and channels. This choice has little impact on the final combined cross sections and is chosen as such as it yields a slightly more conservative total uncertainty for the combined results. The uncertainties from the jet energy scale, the jet energy resolution, \met\ and the \Zjets\ background contribution are treated as fully correlated between all bins and are excluded from the minimisation procedure to avoid numerical instabilities due to the statistical components in these uncertainties. For the combined results, each of these uncertainties is taken as the weighted average of the corresponding uncertainty on the electron and muon measurements, where the weights are the sum in quadrature of all the uncorrelated uncertainties that enter in the combination.

\section{Systematic uncertainties}
\label{sec:sys}

The dominant sources of systematic uncertainties in the cross-section measurements for both the electron and muon channels are the uncertainties in the jet energy 
scale (JES) and at high jet multiplicities the uncertainties on the \ttbar~background estimates. 

Uncertainties in the JES are determined from a combination 
of methods based on simulations and in situ techniques~\cite{Aad:2014bia} and are propagated through the
analysis using 14 independent components, which are fully correlated in jet \pt. These components account for 
uncertainties on the different in situ measurements which form the jet calibration, on the jet flavour and on the impact of pile-up and close-by jets. 
The JES uncertainty varies as a function of jet \pt~and $\eta$ and is less than 2.5\% in the central regions for jets with a \pt\ between 60 \GeV\ and 800 \GeV.
To estimate the impact of the JES uncertainty, jet energies in the simulated events are coherently shifted by the JES uncertainty and the 
missing transverse momentum is recomputed. The full analysis, including re-evaluation of the data-driven background estimates, is repeated with these variations and the cross sections are recomputed;  the change in the cross section is taken as the systematic uncertainty. This method of propagating the uncertainties is also used for most other uncertainties described below. The impact of the JES uncertainties on the cross section for both channels ranges from 9\% for $\njs \ge 1$ to 30\% for $\njs \ge 5$. The uncertainty on the cross section due to the JES for the electron channel is larger because the \Zee\ background is also affected by this uncertainty.

The uncertainty on the jet energy resolution (JER), derived from a comparison of the resolution obtained in data and in simulated dijet events, is propagated into the final 
cross section by smearing the energies of the simulated jets~\cite{Aad:2012ag}. This uncertainty, which is approximately 10\% of the jet energy resolution, results in a 5--20\% uncertainty on the cross sections and is applied symmetrically. 

The uncertainty on the electron and muon selection includes uncertainties on the electron energy or muon momentum scale and resolution, as well as uncertainties on the scale factors applied to the simulations in order to reproduce for electrons or muons the trigger, reconstruction and identification efficiencies measured in the data. The 
lepton energy or momentum scale corrections are obtained from a comparison of the \Zg\ boson invariant mass distribution 
between data and simulations, while the uncertainties on the scale factors are derived from a comparison of tag-and-probe results in data and simulations~\cite{Aad:2011mk,Aad:2014zya}. The overall uncertainty on the cross section is approximately 1--4\%, where the dominant electron uncertainties come from the electron energy scale and identification and the dominant muon uncertainty comes from the trigger.

A residual uncertainty on the \met\ is estimated by scaling the energies of energy clusters in the calorimeters which are not associated with a jet or an electron~\cite{Aad:2012re}.  The resulting uncertainty on the cross section is less than 2\%.

An additional source of uncertainty is a potential bias in the control-sample selection from which multijet templates are extracted. The size of the effect is determined by varying the individual isolation requirements and in the electron channel varying the identification definition, both of which affect the shape of the kinematic distributions of the control sample. To account for shape differences in the low \met\ region, the nominal fit range for the multijet background is varied. The signal template is alternatively modelled by \she\ instead of \alp. 
In addition, for the signal template the uncertainty in the $W/Z$ production cross sections is taken as 5\%~\cite{incW35}. 
The statistical uncertainty on the template normalisation factor from the fit is also included.
The resulting uncertainty on the cross section is 1\% for low jet multiplicities to 25\% at high multiplicities and is dominated by uncertainties in the template shape. 

The dominant uncertainty on the estimate of \ttbar\ background is the statistical uncertainty from the data-driven estimate, which is 6\% on the number of \ttbar\ events for $\njs \ge 3$ to 15\% for $\njs \ge 6$. To estimate the effect due to the subtraction of \Wg+heavy-flavour contamination in the \ttbar\ template, the $W+c$ cross section and the combined $W+c\bar{c}$ and $W+b\bar{b}$ cross sections are varied by factors of 1.3 and 0.9 respectively. These factors are obtained from fits to the selected data in two control 
regions, which have the jet requirements of one or two jets and at least one $b$-tagged jet; in these regions \Wg+heavy flavour events dominate. This uncertainty, which is 3\% of the number of \ttbar\ events for $\njs \ge 3$, is largest at lower jet multiplicities, where the contribution from \Wg+heavy flavour is most significant. Other small uncertainties include uncertainties on the $b$-tagging efficiencies and uncertainties on the bias in the \ttbar\ distributions when applying $b$-tagging. The uncertainty on the number of \ttbar\ events is roughly the same for the electron and muon channels. However, since there are fewer \Wen\ events passing the selection, the relative overall uncertainty on the cross section is larger in the electron channel. The total uncertainty on the cross section for $\njs \ge 4$ due to the estimate of the \ttbar\ background is roughly 10\%. For $\njs \le 2$, where simulations are used to estimate the \ttbar\ background, the uncertainty on  the \ttbar\ cross section is taken to be 6\% as described in Ref.~\cite{Aad:2013cea}.

An uncertainty on the integrated luminosity of 1.8\%~\cite{Aad:2013ucp} is applied to the signal normalisation as well as to all background contributions which are estimated using simulations. 

The uncertainty on the unfolding from the limited number of events in the simulations is estimated using pseudo-experiements. The systematic uncertainties on the unfolding due to modelling in the simulations are estimated by using an alternative set of
\alp\ samples with different parameter values; 
the MLM matching procedure~\cite{Mangano:2002ea} used to remove the double counting between partons
generated from the matrix element calculation and partons from the parton shower
uses a matching cone of size $\Delta R=0.4$ for matrix  element partons of $\pt > 20$~\GeV.
To determine how the arbitrary choice of this cone size and the matching \pt\ scale impacts
the unfolded results, samples where these parameters are varied are used in the
unfolding procedure. In addition, to account for the impact 
of changing the amount of radiation emitted from hard partons, Monte Carlo samples are generated with the renormalisation and 
factorisation scales set to half or twice their nominal value of 
$\sqrt{ m_{W}^2+{\pt}_{W}^2}$. The overall uncertainty on the unfolding procedure
ranges between $0.2 \%$ and $1.7 \%$ over all jet multiplicities.

The systematic uncertainties on the cross-section measurement after unfolding are summarised in Table~\ref{tab:systematics} for both the electron and muon channels and all jet multiplicities. The systematic uncertainties are symmetrised by taking the average value of the up and down variations.

\section{Theoretical predictions}
\label{sec:Theory}

\begin{table*}[htb]
\begin{center}
\begin{tabular}{lcccll} 
%\hline 
\toprule 
 Program & \multicolumn{3}{c}{Max. number of partons at} & Parton/Particle & Distributions  \\ 
  & approx. NNLO & NLO  & LO & level & shown \\
  & ($\alpha_s^{\njs+2})$ & ($\alpha_s^{\njs+1})$ & ($\alpha_s^{\njs})$ &  & \\
  \hline
LoopSim & 1 & 2 & 3 & parton level  &  Leading jet \pt\ and \htj \\
 &  &  & & with corrections & for \Wj \\ [0.25cm] 
\bhs & -- & 5  & 6 & parton level & All \\
 &  & & &  with corrections &  \\ [0.25cm] 
\bhs & 1 & 2  & 3 & parton level  & Leading jet \pt\ and \htj \\
exclusive sums & &  & & with corrections & for \Wj \\ [0.25cm] 
HEJ & \multicolumn{3}{c}{all orders, resummation} & parton level & All \\ 
 &  &  &  & & for $\Wg\,\texttt{+}\mgeq \mathrm{2,\, 3,\, 4\, jets}$  \\ [0.25cm] 
\meps & -- & 2  & 4 & particle level & All \\ [0.25cm] 
\alp & -- & -- & 5  & particle level & All  \\ [0.25cm] 
\she & -- & --  & 4 & particle level & All \\ [0.25cm] 
 \bottomrule
%\hline
\end{tabular}
\caption{Summary of theoretical predictions, including the maximum number of partons at each order in $\alpha_s$, whether or not the results are shown at parton or particle level and the distributions for which they are shown. \label{tab:theory}}
\end{center}
\end{table*}

The measured cross sections for \Wjets\ production are compared to a number of theoretical predictions at both LO and NLO in perturbative QCD, which are summarised in Table~\ref{tab:theory}. The theory predictions are computed in the same phase space in which the measurement is
performed, defined in Sect.~\ref{sec:Unfolding}. The predicted cross sections are multiplied by the branching ratio, $\textrm{Br} (W \rightarrow \ell \nu)$, where $\ell = e,\, \mu$, to compare to the data.

The leading-order predictions shown here include \alp, which is interfaced to HERWIG for showering, \she\, which implements its own parton showering model, and HEJ~\cite{Andersen:2012gk,Andersen:2009nu}, which provides parton-level predictions for \Wjj. \alp\ and \she\ use leading-order matrix element information for predictions of \Wjets\ production and use the MLM~\cite{Mangano:2002ea} and CKKW~\cite{Catani:2001cc} matching schemes, respectively, in order to remove any double counting between the matrix element and parton shower calculations. \alp\ provides predictions with up to five additional partons from the matrix element in the final state while \she\ includes up to four partons.  HEJ is based on a perturbative calculation which gives an approximation to the hard-scattering matrix element for jet multiplicities of two or greater and to all orders in the strong coupling constant, $\alpha_s$.  The approximation becomes exact in the limit of large rapidity separation between partons, also known as the high-energy limit. The resulting formalism is incorporated in a fully exclusive Monte Carlo event generator, from which the predictions shown in this paper are derived. The HEJ results are presented only at the parton level, as the relevant hadronisation corrections are not available, and only for distributions with up to four jets, as the generator version used here is not expected to correctly describe higher multiplicities.

The next-to-leading order predictions at parton level are obtained from \bhs~\cite{Berger:2009ep,Berger:2010zx,Bern:2013gka}, for inclusive \Wnjh\ production, where $n$ ranges from zero to five. The \bh\ program provides the virtual matrix element corrections while \she\ calculates the tree-level diagrams and provides the phase-space integration. 
The \bhs matrix elements are also used in the exclusive sums approach~\cite{AlcarazMaestre:2012vp}, in which NLO information from different jet multiplicities, in this case from $\Wg\,\texttt{+}\, n$ and \Wno\ jets,\footnote{An inclusive NLO prediction for \Wjh\ production explicitly includes (leading-order) corrections from \Wjj, and implicitly, through DGLAP evolution~\cite{Gribov:1972ri, Dokshitzer:1977sg, Altarelli:1977zs}, the effects of additional (collinear) gluon radiation. So in this sense, the calculation includes the effects of additional jets beyond the two included explicitly from the matrix element information.} is utilised. Although not strictly rigorous,\footnote{For example, only the term of order $\alpha_s$ in the strong coupling expansion of the Sudakov form factor expression is used. For a formalism such as MEPS@NLO, as introduced later in the text, the full Sudakov suppression for all jet multiplicities is present.} this approach allows for additional contributions to \Wnjh\ cross sections from higher multiplicity final states than is possible with a normal inclusive prediction. Such contributions can be important when new sub-processes at higher jet multiplicities result in substantial contributions to the cross section.  In practice, these contributions  are most important for predictions involving \Wj. By including such contributions, better agreement between theory and data, as well as smaller theoretical uncertainties, is obtained for several kinematic distributions~\cite{Aad:2012en}.

The next-to-leading order predictions at  particle level are obtained from MEPS@NLO~\cite{Gleisberg:2008ta,Hoeche:2012yf}, which utilises the virtual matrix elements for \Wjeh\ and \Wjjeh\ production determined from \bh, merged with leading-order matrix element information from \Wg\ events with up to four jets. Each final state is then matched to a parton shower and hadronised using \she.  
MEPS@NLO represents a rigorous method of combining NLO + LO matrix element information from a number of different jet multiplicities to produce an exclusive final state at the hadron level. 

Although an NNLO calculation for the production of \Wj\ is not yet available, the LoopSim technique~\cite{Rubin:2010xp} allows the merging of NLO samples of different jet multiplicities in order to obtain approximate NNLO predictions. The LoopSim method makes use of existing virtual matrix elements in the merged samples (here the \Wjeh\ and \Wjjeh\ one-loop virtual matrix elements from MCFM), and where not present, determines exactly the singular terms of the loop diagrams, which, by construction, match precisely the corresponding singular terms of the real diagrams with one extra parton.  The approximate NNLO cross section differs from the complete NNLO cross section only by the constant, i.e. non-divergent parts of the two-loop NNLO terms.
The method is expected to provide predictions close to true NNLO results when the cross sections are dominated by large contributions associated with new scattering topologies that appear at NLO or beyond.

All predictions use CT10 PDFs~\cite{Lai:2010vv}, except for \alp, which uses \cts\ PDFs. The PDF uncertainty is calculated using the CT10 eigenvectors. Since these correspond to a 90\% confidence-level, the resulting uncertainty is scaled down by a factor of 1.645 in order to obtain a one-standard-deviation uncertainty.   The uncertainty due to the value of  $\alpha_s(m_Z)$ is determined by varying the value of $\alpha_s(m_Z)$ by $\pm0.0012$ around the central value of 0.118~\cite{Ball:2012wy}. 

The sensitivity of the theory predictions to higher-order corrections is determined by independently varying the renormalisation and factorisation scales by a factor of two around the central value of $\htj/2$, making sure that the renormalisation and factorisation scales do not differ from each other by more than a factor of two. 

In the following comparisons, the predictions from \bhs\ (both the standard and exclusive sums versions) have uncertainty bands determined by varying the renormalisation and factorisation scales added in quadrature with the 68\% confidence-level uncertainties of the CT10 PDF error set, the $\alpha_s(m_Z)$  uncertainty and the uncertainties from the non-perturbative corrections described below. At low transverse momenta, the PDF+$\alpha_s$ uncertainties and the scale uncertainties are of the same size, with the scale uncertainties increasing in importance as the transverse momentum of the observable increases. The LoopSim predictions have an error band determined by varying the central scale up and down by a factor of two. The HEJ prediction error bands include the 68\% confidence level uncertainties from CT10, along with a variation of the renormalisation and factorisation scales. The \alp, \she\ and MEPS@NLO predictions are shown with the statistical uncertainties related to the size of the generated sample. Although not applied here, the theory uncertainties for \she\ and \alp\ are much larger, as expected from leading-order QCD predictions, while the theory uncertainties for MEPS@NLO for one- and two- jet multiplicities are similar in magnitude to those from \bhs.

\subsection{Non-perturbative and QED final-state radiation corrections}
\label{sec:nonpert}

For comparison to the data, non-perturbative corrections are applied to the parton-level predictions from \bhs\ and LoopSim. These corrections take into account the effects of  hadronisation and of the underlying event and transform the theoretical predictions from the parton level to the particle level. 

The impact of the underlying event tends to add energy to each jet and create additional soft jets while the hadronisation tends to subtract energy from each jet to account for non-perturbative fragmentation effects.  The two effects are thus in opposite directions and mostly cancel each other, leading to a small residual correction. This correction is roughly 10\% of the cross section at low transverse momentum and becomes smaller at higher energies.

The corrections from the parton level to particle level are determined for the \Wjets\ events by making use of \alp\ simulations showered with \her\ and generated with and without the underlying event and with and without non-perturbative fragmentation. The underlying event corrections are calculated using the bin-by-bin ratio of the distributions with the underlying event turned on and off. In a similar manner, the hadronisation correction is computed as the bin-by-bin ratio of particle-level  to parton-level jets. 

The systematic uncertainty on the non-perturbative corrections is determined by calculating the corrections using \alp\ simulations showered with \pyt\ using the \per\ tune. The uncertainty is computed as the difference between the non-perturbative corrections as determined from the two samples. The uncertainty is taken as symmetric around the value of the nominal corrections. 

Comparisons to the data are performed using dressed leptons as described in Sect.~\ref{sec:Unfolding}. To correct parton-level theoretical predictions for QED final-state radiation, a bin-by-bin correction is derived from \alp\ samples for each of the distributions of the measured variables. This is roughly a constant value of 0.99 for most jet multiplicities and for large jet momenta. A systematic uncertainty is determined by comparing the nominal results to those obtained using \she\ samples. The uncertainty is taken as being symmetric and is approximately 0.01 around the nominal values.

\newcommand{\plotsetOneNjet}[4]{
\begin{figure*} \centering
\includegraphics[width=0.80\textwidth]{plot_xsec_paper_#1_#2_w_comb_norm} 
\caption{Cross section for the production of \Wjets\ as a function of the #3.  For the data, the statistical uncertainties are shown by the vertical bars, and the combined statistical and systematic uncertainties are shown by the black-hashed regions. The data are compared to predictions from \bhs, \alp, \she\ and \meps. The left-hand plot shows the differential cross sections and the right-hand plot show the ratios of the predictions to the data. The theoretical uncertainties on the predictions are described in Sect.~\ref{sec:Theory}.}
\label{#4}
\end{figure*}
}

\newcommand{\plotsetHEJNjet}[4]{
\begin{figure*} \centering
\includegraphics[width=0.80\textwidth]{plot_xsec_paper_#1_#2_w_comb_norm} 
\caption{Cross section for the production of \Wjets\ as a function of the #3. For the data, the statistical uncertainties are shown by the vertical bars, and the combined statistical and systematic uncertainties are shown by the black-hashed regions. The data are compared to predictions from \bhs, HEJ, \alp, \she\ and \meps. The left-hand plot shows the differential cross sections and the right-hand plot shows the ratios of the predictions to the data. The theoretical uncertainties on the predictions are described in Sect.~\ref{sec:Theory}.}
\label{#4}
\end{figure*}
}

\newcommand{\plotsetOne}[4]{
\begin{figure*} \centering
\includegraphics[width=0.80\textwidth]{plot_xsec_paper_#1_#2_w_comb_norm} 
\caption{Cross section for the production of \Wjets\ as a function of the #3.  For the data, the statistical uncertainties are shown by the vertical bars, and the combined statistical and systematic uncertainties are shown by the black-hashed regions. The data are compared to predictions from \bhs, \alp, \she\ and \meps. The left-hand plot shows the differential cross sections and the right-hand plot shows the ratios of the predictions to the data. As described in Sect.~\ref{sec:results-njet}, the theoretical predictions have been scaled in order to compare the shapes of the distributions. The theoretical uncertainties, which differ for the various predictions, are described in Sect.~\ref{sec:Theory}. }
\label{#4}
\end{figure*}
}

\newcommand{\plotsetHEJBSEx}[4]{
\begin{figure*} \centering
\includegraphics[width=0.80\textwidth]{plot_xsec_paper_#1_#2_w_comb_norm} 
\caption{Cross section for the production of \Wjets\ as a function of the #3. For the data, the statistical uncertainties are shown by the vertical bars, and the combined statistical and systematic uncertainties are shown by the black-hashed regions. The data are compared to predictions from \bhs, \bhs\ including the exclusive summing, HEJ, \alp, \she\ and \meps. BH+S is an abbreviation for \bhs . The left-hand plot shows the differential cross sections and the right-hand plot shows the ratios of the predictions to the data. As described in Sect.~\ref{sec:results-njet}, the theoretical predictions have been scaled in order to compare the shapes of the distributions. The theoretical uncertainties, which differ for the various predictions, are described in Sect.~\ref{sec:Theory}. }
\label{#4}
\end{figure*}
}

\newcommand{\plotsetEx}[4]{
\begin{figure*} \centering
\includegraphics[width=0.80\textwidth]{plot_xsec_paper_#1_#2_w_comb_norm} 
\caption{Cross section for the production of \Wjets\ as a function of the #3. For the data, the statistical uncertainties are shown by the vertical bars, and the combined statistical and systematic uncertainties are shown by the black-hashed regions. The data are compared to predictions from \bhs, \bhs\ including the exclusive summing, \alp, \she\ and \meps. BH+S is an abbreviation for \bhs . The left-hand plot shows the differential cross sections and the right-hand plot shows the ratios of the predictions to the data. As described in Sect.~\ref{sec:results-njet}, the theoretical predictions have been scaled in order to compare the shapes of the distributions. The theoretical uncertainties, which differ for the various predictions, are described in Sect.~\ref{sec:Theory}. }
\label{#4}
\end{figure*}
}

\newcommand{\plotsetHEJ}[4]{
\begin{figure*} \centering
\includegraphics[width=0.80\textwidth]{plot_xsec_paper_#1_#2_w_comb_norm} 
\caption{Cross section for the production of \Wjets\ as a function of the #3. For the data, the statistical uncertainties are shown by the vertical bars, and the combined statistical and systematic uncertainties are shown by the black-hashed regions. The data are compared to predictions from \bhs, HEJ, \alp, \she\ and \meps. The left-hand plot shows the differential cross sections and the right-hand plot shows the ratios of the predictions to the data. As described in Sect.~\ref{sec:results-njet}, the theoretical predictions have been scaled in order to compare the shapes of the distributions. The theoretical uncertainties, which differ for the various predictions, are described in Sect.~\ref{sec:Theory}. }
\label{#4}
\end{figure*}
}

\newcommand{\plotsetHEJEx}[4]{
\begin{figure*} \centering
\includegraphics[width=0.80\textwidth]{plot_xsec_paper_#1_#2_w_comb_norm} 
\caption{Cross section for the production of \Wjets\ as a function of the #3. For the data, the statistical uncertainties are shown by the vertical bars, and the combined statistical and systematic uncertainties are shown by the black-hashed regions. The data are compared to predictions from \bhs, \bhs\ including the exclusive summing, HEJ, \alp, \she\ and \meps. BH+S is an abbreviation for \bhs . The left-hand plot shows the differential cross sections and the right-hand plot shows the ratios of the predictions to the data. As described in Sect.~\ref{sec:results-njet}, the theoretical predictions have been scaled in order to compare the shapes of the distributions. The theoretical uncertainties, which differ for the various predictions, are described in Sect.~\ref{sec:Theory}. }
\label{#4}
\end{figure*}
}

\newcommand{\plotsetLoopEx}[4]{
\begin{figure*} \centering
\includegraphics[width=0.78\textwidth]{plot_xsec_paper_#1_#2_w_comb_norm} 
\caption{Cross section for the production of \Wjets\ as a function of the #3. For the data, the statistical uncertainties are shown by the vertical bars, and the combined statistical and systematic uncertainties are shown by the black-hashed regions. The data are compared to predictions from \bhs, \bhs\ including the exclusive summing, LoopSim, \alp, \she\ and \meps. BH+S is an abbreviation for \bhs . The left-hand plot shows the differential cross sections and the right-hand plot shows the ratios of the predictions to the data. As described in Sect.~\ref{sec:results-njet}, the theoretical predictions have been scaled in order to compare the shapes of the distributions. The theoretical uncertainties, which differ for the various predictions, are described in Sect.~\ref{sec:Theory}. }
\label{#4}
\end{figure*}
}

\newcommand{\plotsetNoMEPS}[4]{
\begin{figure*} \centering
\includegraphics[width=0.80\textwidth]{plot_xsec_paper_#1_#2_w_comb_norm} 
\caption{Cross section for the production of \Wjets\ as a function of the #3. For the data, the statistical uncertainties are shown by the vertical bars, and the combined statistical and systematic uncertainties are shown by the black-hashed regions. The data are compared to predictions from \bhs, \alp, and \she. The left-hand plot shows the differential cross sections and the right-hand plot shows the ratios of the predictions to the data. As described in Sect.~\ref{sec:results-njet}, the theoretical predictions have been scaled in order to compare the shapes of the distributions. The theoretical uncertainties, which differ for the various predictions, are described in Sect.~\ref{sec:Theory}. }
\label{#4}
\end{figure*}
}

 \begin{table*}
  \centering 
  \begin{tabular}{cll}
  \toprule
  \njs & $\sigma( W \rightarrow \ell\nu\  \texttt{+}\ \ge \njets)$ [pb] \\
\midrule
%         \multicolumn{2}{c}{  $ (W \rightarrow \ell\nu) / (Z \rightarrow \ell\ell) + \geq 0 $ jets }\\
%\midrule
$\geq0$ & $\lbrack~4.849 \pm 0.001$~(stat.)~$\pm 0.05$~(syst.)~$\pm 0.092$~(lumi.)~$] \times 10^3$ \\
$\geq1$ & $\lbrack~4.938 \pm 0.005$~(stat.)~$\pm 0.43$~(syst.)~$\pm 0.097$~(lumi.)~$] \times 10^2$ \\
$\geq2$ & $\lbrack~1.117 \pm 0.002$~(stat.)~$\pm 0.12$~(syst.)~$\pm 0.023$~(lumi.)~$] \times 10^2$ \\
$\geq3$ & $\lbrack~2.182 \pm 0.010$~(stat.)~$\pm 0.31$~(syst.)~$\pm 0.047$~(lumi.)~$] \times 10^1$ \\
$\geq4$ & $\lbrack~4.241 \pm 0.056$~(stat.)~$\pm 0.88$~(syst.)~$\pm 0.095$~(lumi.)~$] \times 10^0$ \\
$\geq5$ & $\lbrack~0.877 \pm 0.032$~(stat.)~$\pm 0.30$~(syst.)~$\pm 0.020$~(lumi.)~$] \times 10^0$ \\
$\geq6$ & $\lbrack~0.199 \pm 0.019$~(stat.)~$\pm 0.11$~(syst.)~$\pm 0.004$~(lumi.)~$] \times 10^0$ \\
$\geq7$ & $\lbrack~0.410 \pm 0.068$~(stat.)~$\pm 0.31$~(syst.)~$\pm 0.009$~(lumi.)~$] \times 10^{-1}$ \\

\bottomrule
\end{tabular}
\caption{Cross section $\sigma( W \rightarrow \ell\nu\  \texttt{+}\ \ge \njets)$  as a function of inclusive jet multiplicity in the phase space defined in the text.}
  \label{tab:njet_incl_w}
\end{table*}

\begin{table*}  
  \centering 
  \begin{tabular}{c|l}
  \toprule
  \njs & $\sigma(W \rightarrow \ell \nu\ \texttt{+}\ \njets)$  [pb] \\
  \midrule
%  \multicolumn{2}{c}{  $ (W \rightarrow \ell\nu) / (Z \rightarrow \ell\ell) + \geq 0 $ jets }\\
%  \midrule
$ = 0$ & $\lbrack~4.343 \pm 0.001$~(stat.)~$\pm 0.06$~(syst.)~$\pm 0.081$~(lumi.)~$] \times 10^3$ \\
$ = 1$ & $\lbrack~3.807 \pm 0.005$~(stat.)~$\pm 0.32$~(syst.)~$\pm 0.073$~(lumi.)~$] \times 10^2$ \\
$ = 2$ & $\lbrack~8.963 \pm 0.016$~(stat.)~$\pm 0.87$~(syst.)~$\pm 0.179$~(lumi.)~$] \times 10^1$ \\
$ = 3$ & $\lbrack~1.755 \pm 0.009$~(stat.)~$\pm 0.23$~(syst.)~$\pm 0.037$~(lumi.)~$] \times 10^1$ \\
$ = 4$ & $\lbrack~3.374 \pm 0.048$~(stat.)~$\pm 0.61$~(syst.)~$\pm 0.075$~(lumi.)~$] \times 10^0$ \\
$ = 5$ & $\lbrack~0.685 \pm 0.027$~(stat.)~$\pm 0.20$~(syst.)~$\pm 0.016$~(lumi.)~$] \times 10^0$ \\
$ = 6$ & $\lbrack~0.160 \pm 0.018$~(stat.)~$\pm 0.09$~(syst.)~$\pm 0.004$~(lumi.)~$] \times 10^0$ \\
$ = 7$ & $\lbrack~0.286 \pm 0.056$~(stat.)~$\pm 0.24$~(syst.)~$\pm 0.006$~(lumi.)~$] \times 10^{-1}$ \\

\bottomrule
\end{tabular}
\caption{Cross section $\sigma(W \rightarrow \ell \nu\ \texttt{+}\ \njets)$ as a function of exclusive jet multiplicity in the phase space defined in the text.}
  \label{tab:njet_excl_w}
\end{table*}

\section {Cross-section results and comparisons to data}
\label{sec:results}

\subsection{Jet multiplicities}
\label{sec:results-njet}

The cross section for \Wln\ production as functions of the inclusive and exclusive jet multiplicity are shown in Figs.~\ref{fig:res-njet} and~\ref{fig:res-njet-excl} and also listed in Tables~\ref{tab:njet_incl_w} and~\ref{tab:njet_excl_w} respectively. In these figures and all following figures, the cross sections are shown for the combined fiducial phase space listed in Table~\ref{tab:selection}. The data are in good agreement with the predictions from \bhs\ for all jet multiplicities up to five jets; above this the experimental uncertainties become large. The \meps\ and HEJ predictions also describe the jet multiplicity cross sections with a similar level of agreement. The \alp\ and \she\ predictions show different trends for jet multiplicities greater than four jets; however, both are in agreement with the data within the experimental systematic uncertainties.

In the following figures, the differential cross sections for the theoretical predictions have been scaled to the measured \Wjets\ cross section in the corresponding jet multiplicity bin shown in Figs.~\ref{fig:res-njet} and~\ref{fig:res-njet-excl} for inclusive and exclusive cross sections respectively, so that the shapes of the distributions can be compared. The factors applied to the theory predictions are summarised in Appendix~\ref{app-aa}. The cross sections for all distributions shown in the paper are available in HepData.\footnote{http://hepdata.cedar.ac.uk/.}

\plotsetHEJNjet{number_of}{jets_incl}{inclusive jet multiplicity}{fig:res-njet}
\plotsetHEJNjet{number_of}{jets_excl}{exclusive jet multiplicity}{fig:res-njet-excl}

\subsection{Jet transverse momenta and rapidities}

The differential cross sections as a function of the leading-jet transverse momentum are shown in Fig.~\ref{fig:res-pt-lead1} for the case of \Wj. The fixed-order theory predictions from \bhs (both the standard and exclusive summing versions) and LoopSim each underestimate the data at high transverse momenta by about two standard deviations of the experimental uncertainty. Although in this region significant contributions are expected from higher-order terms from \Wjj, the results from LoopSim and \bhs\ exclusive sums do not show any significant improvement with respect to \bhs\ in the description of the data. The EWK corrections for inclusive \Wj, which are not included in these predictions, have been calculated~\cite{Denner:2009gj,Chiesa:2013yma} and are sizeable and negative at high \pt. Applying these corrections directly to the \bhs\ predictions would result in a larger discrepancy at large jet transverse momenta. The \alp, \she\ and \meps\ predictions are in fair agreement with the data, although \meps\ shows some deviations at low jet \pt.

The differential cross sections as a function of the exclusive leading-jet \pt, where no second jet is present with a transverse momentum greater than 30 \GeV, are shown in  Fig.~\ref{fig:res-pt-lead1-exc}. There is good agreement between the data and the NLO theoretical predictions (within the large statistical uncertainties), as has also been observed for the \Zjets\ measurements~\cite{Aad:2013ysa}. The requirement that a second jet must not be present reduces the size of the higher-order corrections. However, this good agreement between data and NLO theory is counter-intuitive given that for high values of the leading-jet transverse momentum there is a large disparity of scales (the leading-jet transverse momentum compared to the 30 GeV cut), and  in that situation resummation effects are usually important.

The differential cross section as a function of the leading-jet \pt\ is shown in Fig.~\ref{fig:res-pt-lead2} for \Wjj\ and in Fig.~\ref{fig:res-pt-lead3} for \Wjjj. For two or more jets, the \she\ predictions deviate from the data by up to two standard deviations at high values of the jet \pt, while \bhs\ and \meps\ generally agree well. The \alp\ predictions show similar agreement as for one-jet events. For multiplicities of two or more jets, HEJ can make predictions and it predicts a leading-jet cross section with a harder jet spectrum than present in the data, albeit with large (leading-order) scale uncertainties. For three or more jets, all predictions describe the data well. 

The differential cross sections as a function of the second leading-jet \pt\ are shown in Fig.~\ref{fig:res-pt-sec} for \Wjjh\ production. \alp\ and \she\ generally describe the data well, while the \bhs\ predictions lie below the data for jet $\pt > 100 \GeV$. The \meps\ predictions describe the shape of the data best at high transverse momentum within the large uncertainties but have a different shape below 100~\GeV.
Similar to the leading-jet \pt, HEJ predicts a harder spectrum than present in the data. 

The differential cross sections as a function of the third leading-jet transverse momentum are shown in Fig.~\ref{fig:res-pt-third} for \Wjjj. The predictions are in most cases within one standard deviation of the experimental uncertainties. The one exception is \she, which starts to deviate from the data at high values of the jet \pt. 

The differential cross sections as a function of the fourth leading-jet transvaerse momentum are shown in Fig.~\ref{fig:res-pt-fourth} for \Wjjjj. The HEJ predictions provide a better description here compared to that at lower jet multiplicities. With increasing jet multiplicity, it is more likely that the jets have a similar transverse momenta and that the most forward and backward jets have a larger rapidity separation; in this regime the approximations of HEJ work better. Taking into account the experimental uncertainties, \alp\ and \she\ describe the data fairly well but at large values of the jet \pt\ the two predictions have different trends with respect to the data. 
The \bhs\ predictions lie below the data for the entire transverse momentum range; however, the difference is within the experimental uncertainties. The differential cross sections as a function of the fifth leading-jet transverse momentum are shown in Fig.~\ref{fig:res-pt-fifth} for \Wjjjjj\ and the predictions are all within experimental uncertainties.  

The differential cross sections as a function of the leading-jet rapidity are shown in Fig.~\ref{fig:res-rap-lead1} for \Wjh\ events and the second leading-jet rapidity is shown in Fig.~\ref{fig:res-rap-lead2} for \Wjjh\ events. Overall there is good agreement between the predictions and the data.
For \Wjh\ events, the predictions from \meps, \she\ and to a much lesser extent \bhs\ have a tendency to be higher than the data by one standard deviation of the experimental uncertainty at $|y| > 3.5$, while \alp\ provides a better description. For \Wjjh\ events, similar results are observed although the agreement with the data is better. HEJ provides a good description over the full rapidity range. Similar trends are also seen in measurements by the D0 collaboration~\cite{Abazov:2013gpa}: \she\ overestimates the data at high rapidities while \alp\ provides a better description. Although \alp\ uses a leading-order PDF, if the \alp\ predictions are reweighted to the NLO PDF set CT10, there is no change in the level of agreement with data. An examination of the leading and second-leading jets in \she\, at high rapidities indicates that these jets often originate from the parton shower and therefore disagreements between \alp\ and \she\ most likely arise from the difference in parton showering models. The jet rapidities for the higher jet multiplicities are shown in Appendix~\ref{app-a}.

\subsection{Scalar sums}

The differential cross sections as a function of the \htj\ are shown in Fig.~\ref{fig:res-ht-1} for $\njs \ge 1$ and in Fig.~\ref{fig:res-ht-1ex} for $\njs=1$. For both cases, \alp\ and \she\ tend to be higher than the data at $\htj > 600 \GeV$. The predictions from \bhs are lower than the data for $\njs \ge 1$ and in better agreement for exactly one jet. Better agreement with the data is provided by the \bhs\ exclusive sums and LoopSim predictions, while \meps\ agrees well with the data above 200 \GeV. The \bhs\ exclusive sums and LoopSim predictions are similar to each other at high \htj. This is one of the kinematic variables where the importance of subprocesses such as $qq\rightarrow qqÕW$ (dijet production followed by emission of a $W$ boson from one of the quarks) is most important~\cite{Rubin:2010xp}. The influence of such final states is  reduced when the exclusive one-jet cut is applied, and this is exactly where there is better agreement with the \bhs predictions. 

The higher jet multiplicities are shown in Figs.~\ref{fig:res-ht-2}--\ref{fig:res-ht-5}. The data are, in general, in good agreement with the theoretical predictions, especially the  predictions of \bhs, \meps\ and in some cases \alp. Both the HEJ and \she\ predictions tend to be above the data at high \htj\ but the size of the deviations decreases at higher jet multiplicities. The differential cross sections as a function of the \stj, where \stj\ is defined as the summed scalar \pt\ of all the jets in the event, are shown in Appendix~\ref{app-a} and yield similar conclusions, although  agreement of the theory with the data is better at low \stj\ than at low \htj.

\subsection{Jet angular variables}

Figure~\ref{fig:res-dphi} shows the differential cross sections as a function of the difference in the azimuthal angle (\dpjj) and Fig.~\ref{fig:res-dy} shows the differential cross sections as a function of the difference in the rapidity (\dyjj) between the two leading jets in events with at least two jets. The cross sections as a function of the angular separation (\drjj) are shown in Fig.~\ref{fig:res-dr}  and as a function of the dijet invariant mass in Fig.~\ref{fig:res-mass}. These measurements are tests of hard parton radiation at large angles and matrix element/parton shower matching schemes. Jet production in the forward region can also be very sensitive to the tuning of the underlying event contribution.  

The differential cross sections as a function of the \dpjj\ are fairly well modelled by \bhs, HEJ, \alp\ and \she. For predictions of \dyjj, \bhs\ models the data well while \alp\ underestimates the number of events with very large jet separation and the \she\ and \meps\ predictions overestimate the number of events. This is also reflected in the predictions of \drjj\ where both \alp\ and \she\ have different shapes especially at large values of \drjj. \alp\ underestimates the number of jets with large separation whereas \she\ models the large rapidity intervals better but tends to overestimate the number of close-by jets. \bhs\ shows a similar trend as in the predictions for \dyjj\ but is within the experimental uncertainties. For both variables HEJ underestimates the data for jets with large separation. 

The \she\ and \meps\ predictions fail to model well the region with large values of the dijet invariant mass and overestimate the cross sections. In comparison, the \alp\ predictions underestimate the cross section by one standard deviation of experimental uncertainty. \bhs\ also shows indications of underestimating the number of events at high masses. The HEJ predictions provide a good description of the dijet invariant mass.

\plotsetLoopEx{leading_jet_pt}{njetge1}{leading-jet \pt~in $\njs \ge 1$  events}{fig:res-pt-lead1}
\plotsetOne{leading_jet_pt}{njeteq1}{leading-jet \pt~in $\njs = 1$  events}{fig:res-pt-lead1-exc}

\plotsetHEJ{leading_jet_pt}{njetge2}{leading-jet \pt~in $\njs \ge 2$  events}{fig:res-pt-lead2}
\plotsetHEJ{leading_jet_pt}{njetge3}{leading-jet \pt~in $\njs \ge 3$  events}{fig:res-pt-lead3}

\plotsetHEJ{jet2_pt}{njetge2}{second leading-jet \pt~in $\njs \ge 2$  events}{fig:res-pt-sec}
\plotsetHEJ{jet3_pt}{njetge3}{third leading-jet \pt~in $\njs \ge 3$ events}{fig:res-pt-third}
\plotsetHEJ{jet4_pt}{njetge4}{fourth leading-jet \pt~in $\njs \ge 4$  events}{fig:res-pt-fourth}
\plotsetNoMEPS{jet5_pt}{njetge5}{fifth leading-jet \pt~in $\njs \ge 5$ events}{fig:res-pt-fifth}
%\plotsetOne{jet6_pt}{njetge6}{sixth leading-jet \pt~in $\njs \ge 6$ events}{fig:res-pt-sixth}

\plotsetOne{leading_jet_rap}{njetge1}{leading-jet rapidity in $\njs \ge 1$  events}{fig:res-rap-lead1}
\plotsetHEJ{jet2_rap}{njetge2}{second leading-jet rapidity in $\njs \ge 2$ events}{fig:res-rap-lead2}

\clearpage

\plotsetLoopEx{hht}{njetge1}{\htj\ in $\njs \ge 1$ events}{fig:res-ht-1}
\plotsetOne{hht}{njeteq1}{\htj\ in  $\njs = 1$  events}{fig:res-ht-1ex}
\plotsetHEJ{hht}{njetge2}{\htj\ in $\njs \ge 2$ events}{fig:res-ht-2}
\plotsetHEJ{hht}{njeteq2}{\htj\ in  $\njs = 2$ events}{fig:res-ht-2ex}
\plotsetHEJ{hht}{njetge3}{\htj\ in $\njs \ge 3$ events}{fig:res-ht-3}
\plotsetHEJ{hht}{njeteq3}{\htj\ in  $\njs = 3$ events}{fig:res-ht-3ex}
\plotsetHEJ{hht}{njetge4}{\htj\ in $\njs \ge 4$ events}{fig:res-ht-4} 
\plotsetOne{hht}{njetge5}{\htj\ in $\njs \ge 5$ events}{fig:res-ht-5}

\plotsetHEJ{DeltaPhi_jet1_jet2}{njetge2}{difference in the azimuthal angle between the two leading jets in $\njs \ge 2$ events}{fig:res-dphi}
\plotsetHEJ{DeltaRap_jet1_jet2}{njetge2}{difference in the rapidity between the two leading jets in $\njs \ge 2$ events}{fig:res-dy}
\plotsetHEJ{DeltaR_jet1_jet2}{njetge2}{angular separation between the two leading jets in $\njs \ge 2$ events}{fig:res-dr}
\plotsetHEJ{hmj1j2}{njetge2}{dijet invariant mass ($m_{12}$) between the two leading jets in $\njs \ge 2$ events}{fig:res-mass}

\clearpage

\section{Summary}
\label{sec:summary}

In this paper, results are presented for the production of a $W$ boson plus jets, measured in proton--proton collisions at $\sqrt{s}=7\TeV$ with the ATLAS experiment at the LHC. Final states with up to seven jets are measured, with comparisons to precision NLO QCD predictions for up to five jets. With an integrated luminosity of $4.6\,\ifb$, this data set allows an exploration of a large kinematic range, including jet production up to a  transverse momentum of  $1 \TeV$. 

The data are compared to a variety of theoretical predictions, at both leading order and next-to-leading order and the results presented are, with some exceptions, in good agreement. However there is currently no theoretical prediction that is able to provide an accurate description of the data for all measured differential cross sections. Fixed-order predictions, such as \bhs, provide overall a good description of the data, but have greater difficulty  describing variables such as \htj\ or \stj\ in kinematic regions where the dominant production mechanism is dijet production, followed by the emission of a $W$ boson from one of the quarks. Here better agreement is provided by extensions to fixed-order predictions, such as LoopSim or the \bhs\ exclusive sums method, or by formalisms that naturally include higher-order matrix element information within a Monte Carlo parton shower formalism, such as \meps. The predictions of HEJ agree better with the data in regions where there is a large jet multiplicity and/or the jets tend to be separated by a wider rapidity interval. The leading-order matrix element calculations of \alp\ and \she\ provide a good description of the data for most differential cross sections but fail to describe jets with large rapidities and large angular separations. 

The data presented in this paper, for $W$ production in association with jets, will allow a better quantitative understanding of perturbative QCD as well as future comparisons to predictions which include EWK corrections.

\section*{Acknowledgements}
\label{ack}

We thank CERN for the very successful operation of the LHC, as well as the
support staff from our institutions without whom ATLAS could not be
operated efficiently.

We acknowledge the support of ANPCyT, Argentina; YerPhI, Armenia; ARC,
Australia; BMWFW and FWF, Austria; ANAS, Azerbaijan; SSTC, Belarus; CNPq and FAPESP,
Brazil; NSERC, NRC and CFI, Canada; CERN; CONICYT, Chile; CAS, MOST and NSFC,
China; COLCIENCIAS, Colombia; MSMT CR, MPO CR and VSC CR, Czech Republic;
DNRF, DNSRC and Lundbeck Foundation, Denmark; EPLANET, ERC and NSRF, European Union;
IN2P3-CNRS, CEA-DSM/IRFU, France; GNSF, Georgia; BMBF, DFG, HGF, MPG and AvH
Foundation, Germany; GSRT and NSRF, Greece; RGC, Hong Kong SAR, China; ISF, MINERVA, GIF, I-CORE and Benoziyo Center, Israel; INFN, Italy; MEXT and JSPS, Japan; CNRST, Morocco; FOM and NWO, Netherlands; BRF and RCN, Norway; MNiSW and NCN, Poland; GRICES and FCT, Portugal; MNE/IFA, Romania; MES of Russia and ROSATOM, Russian Federation; JINR; MSTD,
Serbia; MSSR, Slovakia; ARRS and MIZ\v{S}, Slovenia; DST/NRF, South Africa;
MINECO, Spain; SRC and Wallenberg Foundation, Sweden; SER, SNSF and Cantons of
Bern and Geneva, Switzerland; NSC, Taiwan; TAEK, Turkey; STFC, the Royal
Society and Leverhulme Trust, United Kingdom; DOE and NSF, United States of
America.

The crucial computing support from all WLCG partners is acknowledged
gratefully, in particular from CERN and the ATLAS Tier-1 facilities at
TRIUMF (Canada), NDGF (Denmark, Norway, Sweden), CC-IN2P3 (France),
KIT/GridKA (Germany), INFN-CNAF (Italy), NL-T1 (Netherlands), PIC (Spain),
ASGC (Taiwan), RAL (UK) and BNL (USA) and in the Tier-2 facilities
worldwide.

\clearpage
\newpage
%\bibliographystyle{atlasnote}

%EPJC \bibliographystyle{spphys}
\bibliographystyle{atlasBibStyleWoTitle}

\bibliography{wjet2011}

\clearpage
\newpage

\appendix{Appendices}

\section{Scale factors for theoretical predictions}
\label{app-aa}

\begin{table*}[htb]
\begin{center}
\begin{tabular}{lcccccccc} 
%\hline 
\toprule 
$N_{\mbox{jet}}$ & $\ge 1$ &  $= 1$ & $\ge 2$  &  $= 2$  & $\ge 3$  & $= 3$ & $\ge 4$ & $\ge 5$ \\
\midrule
LoopSim & 1.029 & -- & -- & -- & -- & -- & -- & -- \\
\bhs & 0.960 & 0.969 & 1.003 & 1.002 & 1.075 & 1.044 & 1.101 & 1.064 \\
\bhs ex. sum. & 0.960 & -- & -- & -- & -- & -- & -- & -- \\
HEJ & -- & -- & 0.960 & 0.932 & 1.091 & 1.123 & 0.968 & -- \\
\meps & 1.099 & 1.105 & 1.094 & 1.095 & 1.103 & 1.094 & 1.146 & 1.183 \\
\alp & 0.940 & 0.945 & 0.936 & 0.935 & 0.946 & 0.946 & 0.960 & 0.856 \\
\she & 0.925 & 0.939 & 0.892 & 0.880 & 0.948 & 0.919 & 1.074 & 1.209\\
\bottomrule
%\hline
\end{tabular}
\caption{Summary of the scale factors applied to the theoretical predictions in the differential cross-section distributions. \label{tab:theory-scaling}}
\end{center}
\end{table*} 

\clearpage

\section{Additional jet-rapidity and \stj\ distributions}
\label{app-a}

\plotsetHEJ{jet3_rap}{njetge3}{third leading jet rapidity in $\njs \ge 3$  events}{fig:res-rap-lead3}
\plotsetHEJ{jet4_rap}{njetge4}{fourth leading jet rapidity in $\njs \ge 4$ events}{fig:res-rap-lead4} 
\plotsetNoMEPS{jet5_rap}{njetge5}{fifth leading jet rapidity in $\njs \ge 5$ events}{fig:res-rap-lead5}

\plotsetEx{hst}{njetge1}{\stj\ in $\nj \ge 1$ events}{fig:res-st-1}
\plotsetHEJ{hst}{njetge2}{\stj\ in $\njs \ge 2$ events}{fig:res-st-2}
\plotsetHEJ{hst}{njeteq2}{\stj\ in  $\njs = 2$ events}{fig:res-st-2ex}
\plotsetHEJ{hst}{njetge3}{\stj\ in $\njs \ge 3$ events}{fig:res-st-3}
\plotsetHEJ{hst}{njeteq3}{\stj\ in  $\njs = 3$ events}{fig:res-st-3ex}
\plotsetHEJ{hst}{njetge4}{\stj\ in $\njs \ge 4$ events}{fig:res-st-4}
\plotsetOne{hst}{njetge5}{\stj\ in $\njs \ge 5$ events}{fig:res-st-5}

\clearpage

\clearpage

\onecolumn
% ATLAS Collaboration author list
% Data extracted on 26-Jan-2015 for paper reference STDM-2012-24

\begin{flushleft}
{\Large The ATLAS Collaboration}

\bigskip

G.~Aad$^{\rm 84}$,
B.~Abbott$^{\rm 112}$,
J.~Abdallah$^{\rm 152}$,
S.~Abdel~Khalek$^{\rm 116}$,
O.~Abdinov$^{\rm 11}$,
R.~Aben$^{\rm 106}$,
B.~Abi$^{\rm 113}$,
M.~Abolins$^{\rm 89}$,
O.S.~AbouZeid$^{\rm 159}$,
H.~Abramowicz$^{\rm 154}$
% ...
\bigskip
\\
% ...
$^{11}$ Institute of Physics, Azerbaijan Academy of Sciences, Baku, Azerbaijan\\
% ...
$^{84}$ CPPM, Aix-Marseille Universit{\'e} and CNRS/IN2P3, Marseille, France\\
% ...
$^{89}$ Department of Physics and Astronomy, Michigan State University, East Lansing MI, United States of America\\
% ...
$^{106}$ Nikhef National Institute for Subatomic Physics and University of Amsterdam, Amsterdam, Netherlands\\
% ...
$^{112}$ Homer L. Dodge Department of Physics and Astronomy, University of Oklahoma, Norman OK, United States of America\\
$^{113}$ Department of Physics, Oklahoma State University, Stillwater OK, United States of America\\
% ...
$^{116}$ LAL, Universit{\'e} Paris-Sud and CNRS/IN2P3, Orsay, France\\
% ...
$^{152}$ Institute of Physics, Academia Sinica, Taipei, Taiwan\\
% ...
$^{154}$ Raymond and Beverly Sackler School of Physics and Astronomy, Tel Aviv University, Tel Aviv, Israel\\
% ...
$^{159}$ Department of Physics, University of Toronto, Toronto ON, Canada\\
% ...
\end{flushleft}

%\end{linenumbers}

%CONF \input{additional_conf_plots}
%\input{additional_plots}
%\input{additional_tables_wjets}

\end{document}